\documentclass[submission,copyright,creativecommons]{eptcsstyle/eptcs}
\providecommand{\event}{ACCAT 2012} 
\def\titlerunning{Characterizing Van Kampen Squares}
\def\authorrunning{H. K\"{o}nig, U. Wolter, M.L\"{o}we}
\usepackage[english]{babel}
\usepackage{makeidx}
\usepackage{amsmath}
\usepackage{amssymb}
\usepackage[mathcal]{euscript}
\usepackage{varioref}
\usepackage[all]{xy}

\usepackage{float}
\restylefloat{figure}

\usepackage{url}
\usepackage{graphicx}

\usepackage{color}

\newdir{ (}{{}*!/-5pt/@^{(}}    
\newdir{ )}{{}*!/-5pt/@_{(}}    
\newdir{ >}{{}*!/-5pt/@{>}}     

\def\squareforqed{\hbox{\rlap{$\sqcap$}$\sqcup$}}
\def\qed{\ifmmode\squareforqed\else{\unskip\nobreak\hfil
\penalty50\hskip1em\null\nobreak\hfil\squareforqed
\parfillskip=0pt\finalhyphendemerits=0\endgraf}\fi}

\newcommand{\C}{\mathcal C} 			

\newcommand{\CO}[1]{\C\hspace{-0.3ex}\downarrow\hspace{-0.3ex}{#1}} 		
\newcommand{\Ob}[1]{\mathit{Ob}_{#1}} 	
\newcommand{\Mor}[1]{\mathit{Mor}_{#1}} 	

\newcommand{\FunctorF}{\mathcal{F}}

\newcommand{\FunctorT}{\mathcal{T}}

\newcommand{\Mapping}[3]{\xymatrix{{#1} \ar[r]^{{#2}} & {#3}}}
\newcommand{\MonoMapping}[3]{\xymatrix{{#1} \ar@{  >->}[r]^{{#2}} & {#3}}}
\newcommand{\EpiMapping}[3]{\xymatrix{{#1} \ar@{->>}[r]^{{#2}} & {#3}}}
\newcommand{\IsoMapping}[3]{\xymatrix{{#1} \ar@{  >->>}[r]^{{#2}} & {#3}}}
\newcommand{\InclMapping}[3]{\xymatrix{{#1} \ar@{ (->}[r]^{{#2}} & {#3}}}

\newcommand{\myref}[1]{(\ref{#1})}
\renewcommand{\a}{\alpha}

\newcommand{\e}{\varepsilon}

\renewcommand{\o}[1]{\overline{#1}}
\newcommand{\N}{\mathbb{N}}
\newcommand{\la}{\langle}
\newcommand{\ra}{\rangle}

\newtheorem{lemma}{Lemma}
\newtheorem{proposition}[lemma]{Proposition}
\newtheorem{definition}[lemma]{Definition}

\newtheorem{theorem}[lemma]{Theorem}
\newtheorem{example}[lemma]{Example}
\begin{document}
\entrymodifiers={+!!<0pt,\fontdimen22\textfont2>}

\title{Characterizing Van Kampen Squares via Descent Data}
\author{
	Harald K\"{o}nig
	\institute{University of Applied Sciences\\ FHDW Hannover, Germany}
	\email{harald.koenig@fhdw.de}
\and
	Uwe Wolter
	\institute{Department of Informatics\\ University of Bergen, Norway}
	\email{wolter@ii.uib.no}
\and
	Michael L\"{o}we
	\institute{University of Applied Sciences\\ FHDW Hannover, Germany}
	\email{michael.loewe@fhdw.de}
}

\maketitle
\begin{abstract}
Categories in which cocones satisfy certain exactness conditions w.r.t. pullbacks are subject to current research activities in theoretical computer science. Usually, exactness is expressed in terms of properties of the pullback functor associated with the cocone. Even in the case of non-exactness, researchers in model semantics and rewriting theory inquire an elementary characterization of the image of this functor. In this paper we will investigate this question in the special case where the cocone is a cospan, i.e. part of a {\em Van Kampen square}. The use of {\em Descent Data} as the dominant categorical tool yields two main results: A simple condition which characterizes the reachable part of the above mentioned functor in terms of liftings of involved equivalence relations and (as a consequence) a necessary and sufficient condition for a pushout to be a Van Kampen square formulated in a purely algebraic manner.
\end{abstract}

\section{Introduction}\label{section1}
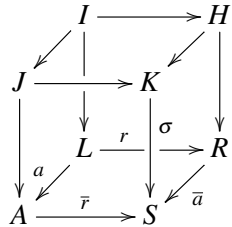
\begin{figure}[!htb]
\[
\xymatrix{
						&{I}\ar[dd]|{\hole}\ar[ld]\ar[rr]		& {}		& {H}\ar[dd]\ar[dl]		\\
	{J}\ar[dd]\ar[rr]	&{} 									& {K}\ar[dd]^(0.33){\sigma}& {} 					\\
						&{L}\ar[rr]^<<<<{r}|{\hole}\ar[ld]_a 	& {} 		& {R} \ar[ld]^{\o{a}}			\\
{A}\ar[rr]^{\o{r}}				& 				 						& {S} 		& {} 	 				\\	
}
\]
\caption{Van Kampen square}\label{Fig-1}
\end{figure}

There is a considerable amount of theoretical work in software engineering and category theory that has frequently encountered the question whether the interplay of pushouts and pullbacks satisfies certain exactness conditions. There is ongoing research in classifying and characterizing categories in which co\-limits and pullbacks are reasonably related. A prominent example are adhesive categories \cite{LS04}, in which pushouts along monomorphisms are {\em Van Kampen squares}. However, this property can be formulated for any commutative square in the bottom of Figure \ref{Fig-1} as follows: The functor $PB$, which takes $\sigma\in\CO{S}$ and maps it to a rear pullback span by pulling back along $\o{a}\circ r = \o{r}\circ a$, has to be an equivalence of categories. 
\par
\xymatrixrowsep{1.0pc} 
\xymatrixcolsep{1.0pc}
If the bottom square is a pushout, the following property is equivalent to this definition: In every commutative cube as in Figure \ref{Fig-1} with two pullbacks as rear faces, the following equivalence holds: The top face is a pushout if and only if the front and right faces are pullbacks \cite{Sobo2004}. 
\xymatrixrowsep{1.6pc} 
\xymatrixcolsep{1.6pc}
\par
The category $SET$ of sets and mappings between them as well as the category $GRAPH$ of graphs\footnote{
	i.e. directed graphs $(V,E;s,t:E\to V)$
} and graph morphisms are adhesive. In many applications (e.g. \cite{Ehrig2006-foagt,Sobo2004}) it is sufficient to infer exactness for pushouts along monomorphisms only. However, it is well-known that already in $SET$ there are many more Van Kampen squares than the ones where one participating morphism is monic. Additionally, several research topics have evolved, where the implications of the Van Kampen property were needed in the case of non-monic $a$ and $r$.
\par
\begin{figure}[!htb]
\[
\xymatrix{
& 	L \ar[rr]^r\ar[dl]_a\ar@/_0.5pc/[dddr]|<<<<<<<<<<{\hole}_>>>>>>>>>>>>{\gamma}		&		& R\ar[dl]_{\o{a}}\ar@/^1pc/[dddl]^\beta	&&
					& 		{I}\ar[dd]|{\hole}^<<<<{\gamma}\ar[ld]_{a'}\ar[rr]^{r'}& {} & {H}\ar[dd]^{\beta}\ar@{-->}[dl]^{?}\\
A\ar[rr]^<<<<<<{\o{r}}\ar@/_1pc/[ddrr]_\tau&	&S\ar@{-->}[dd]^{!}	&										&&						 
		{J}\ar[dd]^{\tau}\ar@{-->}[rr]_<<<<<{?}	&	{} 									& {K}\ar@{-->}[dd]^<<<<{?}		& {} 			 \\
& 		& 			& 																				&&	
					& 		{L}\ar[rr]|{\hole}^<<<<{r}\ar[ld]^{a}								& {} 				& {R} \ar[ld]^{\overline{a}}\\
& 		& SET		& 																				&&	
		{A}\ar[rr]_{\overline{r}}	& 		{} 							& S		 	& {}  \\	
}
\]
\caption{Indexed vs Fibred Amalgamation of $\tau$ and $\beta$ with common part $\gamma$}\label{Fig0a}
\end{figure}
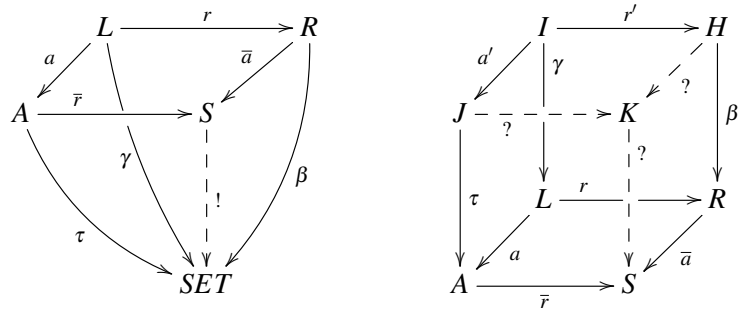

An important example are {\em diagrammatic specifications}\footnote{
	E.g. UML class diagrams or ER diagrams 
} 
in model driven engineering \cite{WD2008_a}: In Figure \ref{Fig0a}, there are specifications $A, L, R$, and $S$ each of which contain (data) types and directed relations between them, i.e they are small graphs. Since the specifications require {\em compositionality} \cite{EGW1998}, it is important to investigate {\em amalgamation}, a simple and natural construction which provides the basis for compositionality. It is a method to uniquely and correctly compose interpretations of parts of an already composed specification. Formulated in indexed semantics this takes the form as shown in the left diagram of Figure \ref{Fig0a}: There is a pushout of specification morphisms as top face together with interpretations $\tau$ and $\beta$ with common part $\gamma$, i.e. $\beta\circ r = \gamma = \tau\circ a$. Here the large graph $SET$ serves as a ''semantic universe''. All arrows in Figure \ref{Fig0a} are graph homomorphisms. A unique and correct amalgamation of interpretations for the indexed case comes quite naturally by constructing the unique mediating arrow from $S$ to $SET$.
\par
\par
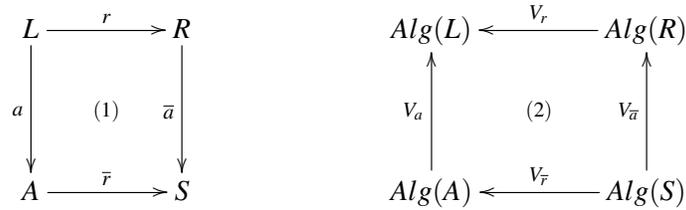
\begin{figure}[!htb]
\[\xymatrix{
	L\ar[rr]^r\ar[dd]_a	&&	R	\ar[dd]_{\o{a}}		&&&	Alg(L)\ar@{}[ddrr]|{(2)}&& 	Alg(R)\ar[ll]_{V_r}													\\
											&&											&&&													&&																							\\
	A\ar[rr]^{\o{r}}		&&	S\ar@{}[uull]|{(1)} &&&	Alg(A)\ar[uu]^{V_a}			&& Alg(S)\ar[uu]^{V_{\o{a}}}\ar[ll]_{V_{\o{r}}} \\
}\] 
\caption{Amalgamation Lemma (Indexed setting)}\label{Fig0b}
\end{figure}

 
There is also a global view on the amalgamation procedure in the indexed setting: Let us denote the category of interpretations of a specification $X$ by $Alg(X)$\footnote{
	We use this abbreviation, because the term ''interpretation'' is often substituted by the term ''algebra''. 
} and let $V$ denote the usual forgetful functor along a specification morphism (e.g.\ the functor $V_r:Alg(R)\to Alg(L)$ is defined by $V_r(\beta) = \beta\circ r$), cf.\ Figure \ref{Fig0b}. The Amalgamation Lemma \cite{EM+85} states that (2) is a pullback in the category $CAT$ of categories if (1) is a pushout of specifications. 
\par 
Note that the ''philosophy'' of semantic universes implies two important facts: On the one hand, elements of a set (objects) can be multiply interpreted (typed) (if e.g. $\gamma(t_1)\cap \gamma(t_2) \not=\emptyset$ for two different nodes $t_1, t_2$ in $L$). On the other hand, one can determine all objects that are $t$-typed by considering $\gamma(t)$.    
\par
But reliable semantics for model-driven structures has to omit the ''philosophy'' of semantic universes, because in software environments each object possesses exactly one type and it should not be possible to determine the set of $t$-typed objects \footnote{
	Consider conformance relations in standards of software engineering (e.g. UML object diagrams or MOF) \cite{rutle2008-diagrammatic}. 
}. This mismatch requires the shift from indexed to fibred semantics \cite{WD2008_a}. In the fibred setting, interpretations are called {\em instances} and are formalized by objects of the slice categories $\CO{A}$, $\CO{L}$, $\CO{R}$, and $\CO{S}$. Forgetful functors are now ''pulling back''-functors (e.g.\ the functor $r^\ast$ which constructs the pullback of $(r,\beta)$, see the right part of Figure \ref{Fig0a}). 
\par 
This raises the question whether the amalgamation procedure smoothly carries over to the fibred setting. I.e. given two instances $\tau\in \CO{A}$ and $\beta\in \CO{R}$ with common part $\gamma$, i.e. $r^\ast\beta = \gamma = a^\ast\tau$, one wants to prove that the syntactical composition (pushout of $a$ and $r$) is reflected on the instance level by a unique construction. The counterpart for correctness is the requirement to obtain an $S$-instance of $\CO{S}$, such that its pullbacks along $\o{a}$ and $\o{r}$ yield $\beta$ and $\tau$, resp, cf.\ Figure \ref{Fig0a}. 

\section{The Reachability Problem}\label{section1,5}
In contrast to indexed amalgamation, there are intrinsic difficulties for the fibred setting, because the given rear pullback span must not be in the image of $PB$. In other words, a reasonable construction on the instance-level fails if and only if the  pullback span is not {\em reachable} by $PB$. This is demonstrated in  
\begin{example}\label{Example1}
In Figure \ref{FigExample1}, objects are denoted $i\hspace{-0.2em}:\hspace{-0.2em}t$, instances map objects to their types. $a$ and $r$ map according to the letters. $i\hspace{-0.2em}:\hspace{-0.2em}t, j\hspace{-0.2em}:\hspace{-0.2em}s\in I$ are connected via dashed lines if $r'(i\hspace{-0.2em}:\hspace{-0.2em}t) = r'(j\hspace{-0.2em}:\hspace{-0.2em}s)$. Dotted lines depict the kernel of $a'$. It can easily be computed that the two rear squares establish a pullback span in $SET$. 
\par 
However, the span is not reachable: On the one hand, pullback complements for the right and the front face with sets over $S$ containing two elements will always yield a non-commutative top face. On the other hand, the pushout on the top face creates a $\CO{S}$-object (the mediator out of the pushout), whose domain is a singleton set. But pulling back this instance along $\o{r}$, $\o{a}$ resp.\ does not yield $\tau$ and $\beta$, resp.
\begin{figure}[!htb]
\begin{center}
\includegraphics[height=2in]{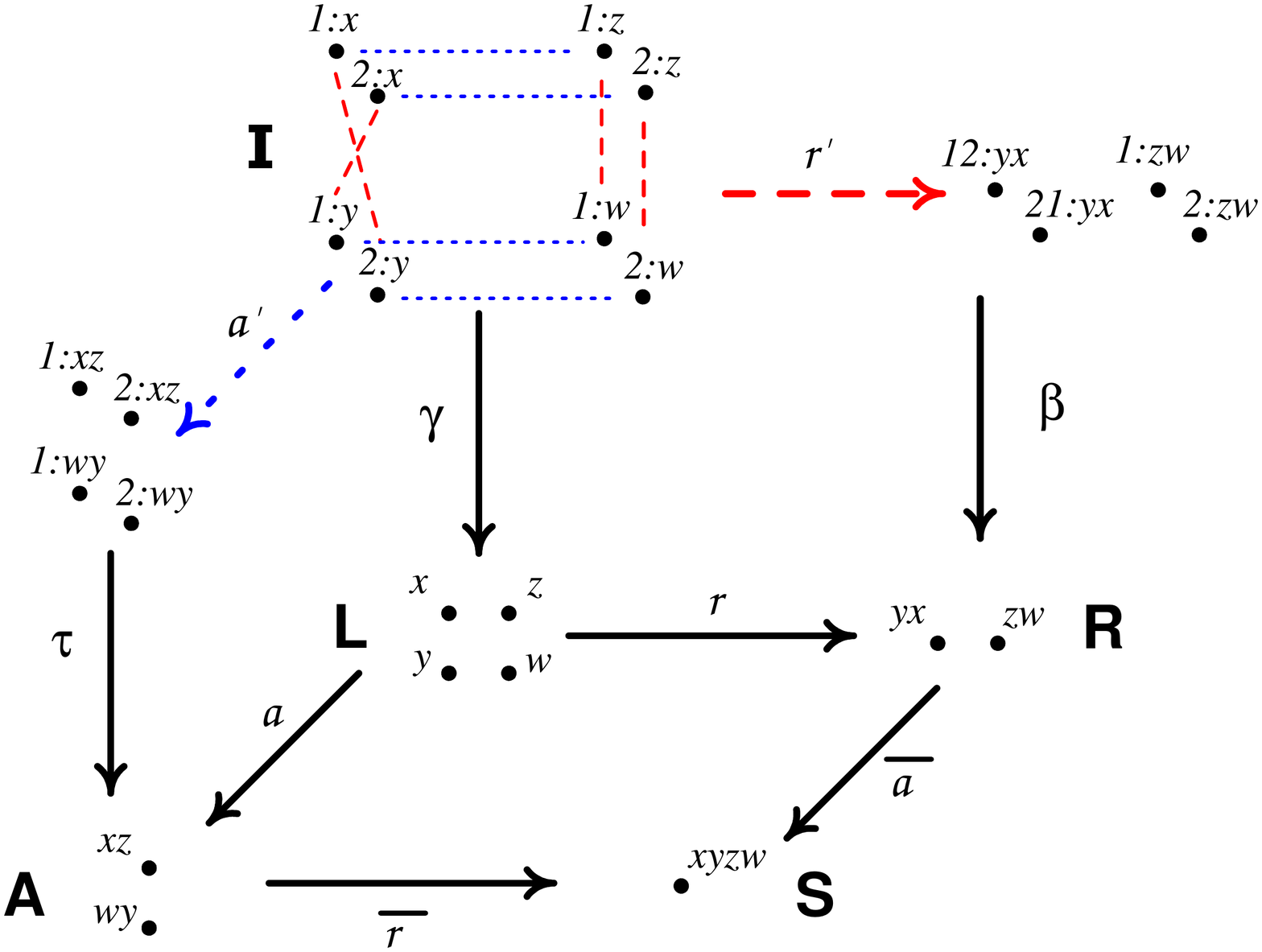}
\caption{Unreachable pullback span}\label{FigExample1}
\end{center}
\end{figure}
\end{example}
These effects can not occur in the indexed setting because multiple typing was allowed. To get rid of multiple typing, the transition from indexed to fibred semantics entails the production of copies. E.g. in the indexed setting, it would be sufficient to let $\gamma$ map each element of $L$ to the set $\{1,2\}$, whereas the fibred view requires to produce $4$ copies of this $2$-element set (yielding the set $I$ in Figure \ref{FigExample1}). It is well-known that indexed categories are related to fibrations via the Grothendieck construction \cite{BaWe1990,WD2007}. However, since the image of this construction is the category of {\em split} fibrations, all produced copies behave in a uniform way as in the next example.
\begin{example}\label{Example2}
In this example fibres are lifted in a uniform way. The pullback span is now reachable. It is isomorphic to $PB(\sigma)$ where $\sigma:\{1\hspace{-0.2em}:\hspace{-0.2em}xyzw, 2\hspace{-0.2em}:\hspace{-0.2em}xyzw\}\to S$. 
\begin{figure}[!htb]
\begin{center}
\includegraphics[height=2in]{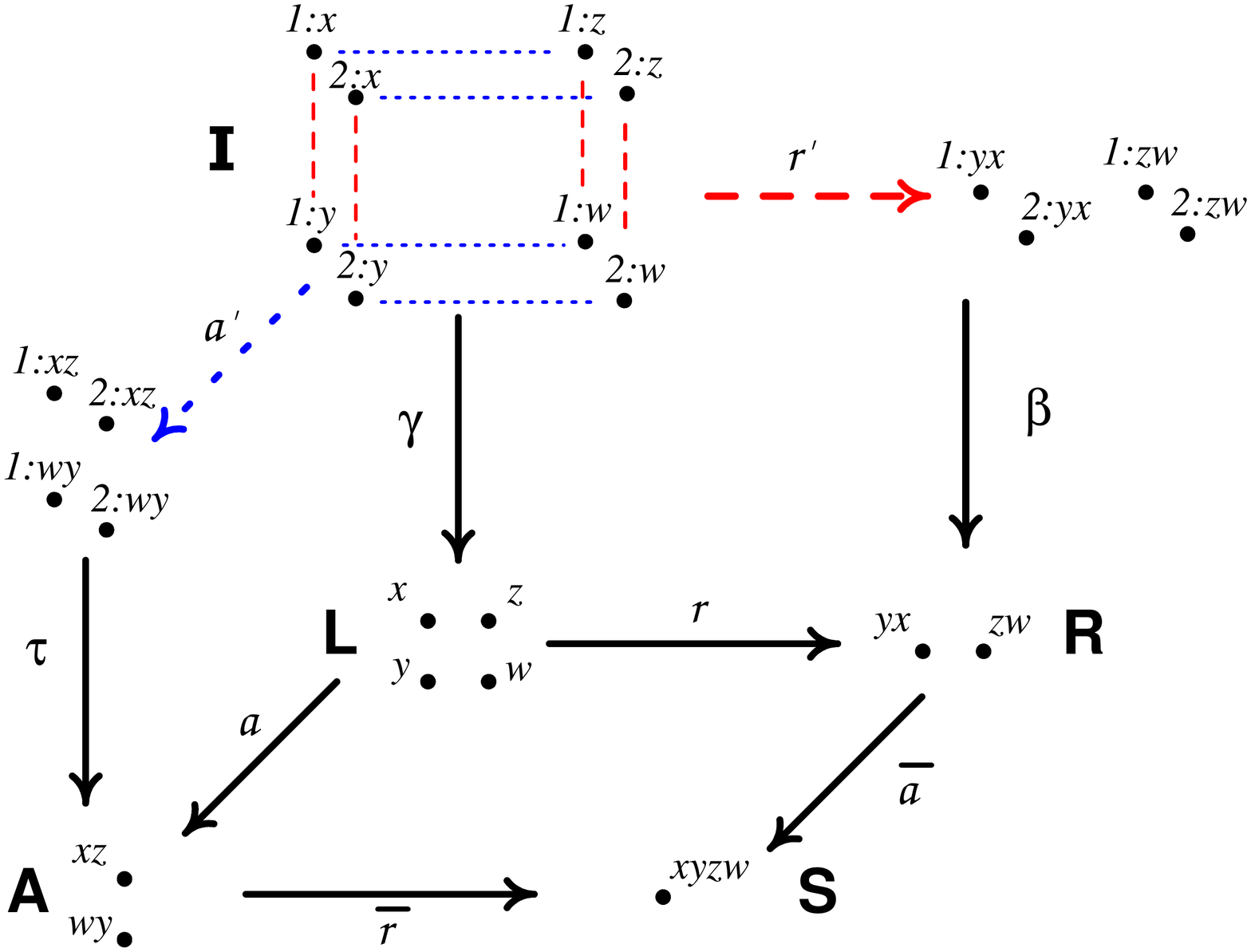}\caption{Reachable pullback span}\label{Fig9.5}
\end{center}
\end{figure} 
\end{example}
But if pullback spans are not results of the Grothendieck construction, we suffer from the enlarged degree of freedom for defining the relationship between fibres, i.e. the equivalence relations of $a'$ and $r'$ may chaotically be intertwined as in Example \ref{Example1}. 
\par 
Of course, fibred amalgamation is successful, if the bottom square in the cube of Figure \ref{Fig0a} would be a Van Kampen square. In this case, one simply has to construct the pushout on top of the cube and can automatically deduce that this produces two pullbacks in front as desired. Then the question arises, how to detect whether a square is a Van Kampen square from properties of $a$ and $r$ only. 
\begin{example}\label{Example3}
In the pushout in Figure \ref{FigExample3} neither $a$ nor $r$ is monic. Hence we cannot infer the Van Kampen property from the fact that $SET$ is an adhesive category. 
\begin{figure}[!htb]
\begin{center}
\includegraphics[height=1.2in]{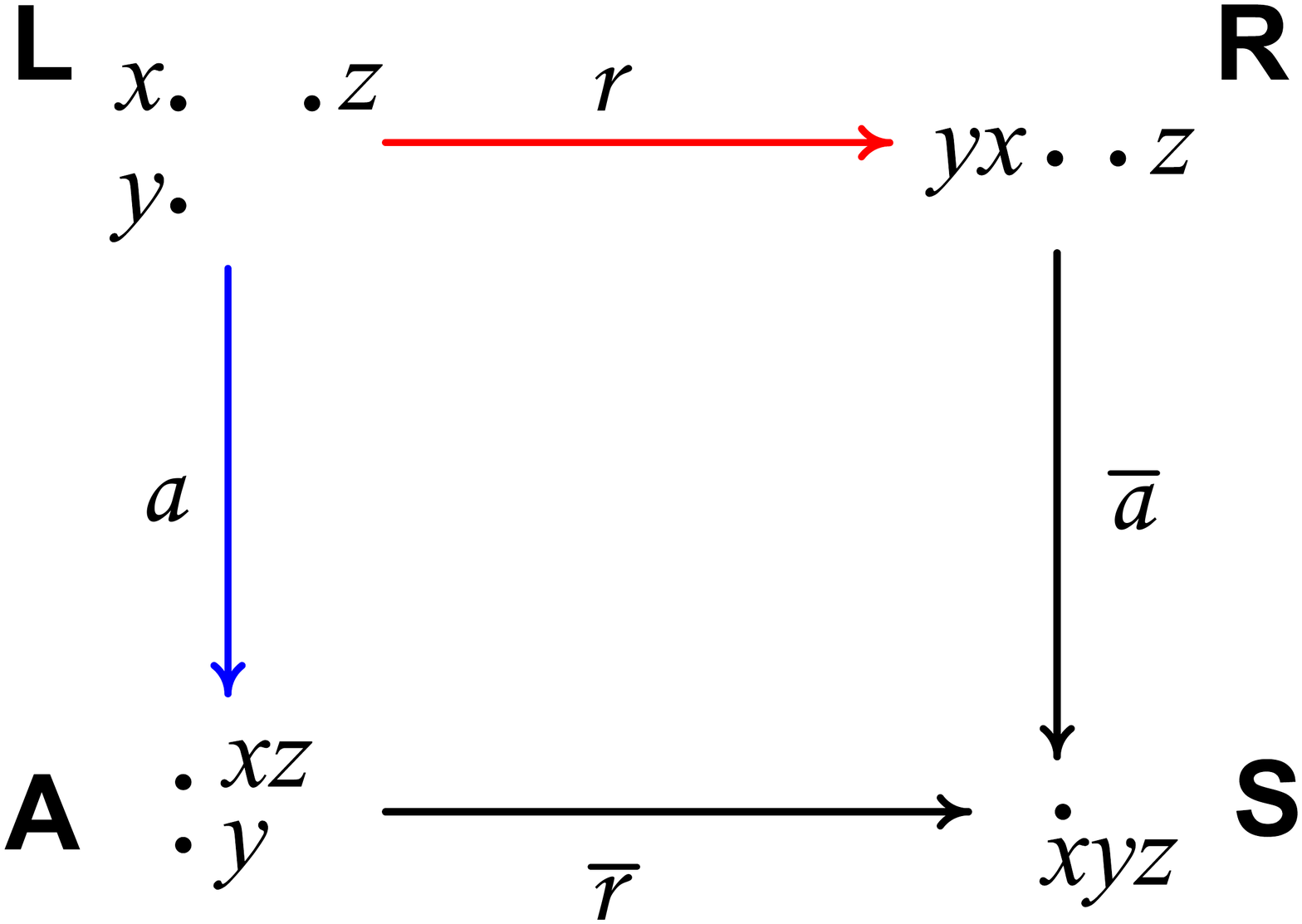}\caption{Van Kampen square?}\label{FigExample3}
\end{center}
\end{figure} 
\end{example}
\begin{description}
\item[{\em Question 1}] {\em Can we find a feasable condition which characterizes reachability in terms of the rear pullback span only (even in the case that the bottom square is not a Van Kampen square)?}	
\item[{\em Question 2}] {\em Can we find a necessary and sufficient condition for a pushout to be a Van Kampen square in terms of the span $(a,r)$ only?}
\end{description}
It became evident that a comprehensive investigation has to be performed in a more abstract categorical environment, see also \cite{L2010}. A good generalization are {\em topoi} \cite{Goldblatt1984-t}, i.e. categories which have finite limits, are cartesian closed, and where the subobject functor is representable\footnote{
	In the sequel, we assume the reader to possess basic understanding of the notion of topos.
}. $SET$ and $GRAPH$ are topoi. Topoi are adhesive \cite{LS06}, i.e.\ pushouts along monomorphisms are Van Kampen squares. Thus the above questions are relevant only for the case where both $a$ and $r$ in Figure \ref{Fig0a} have non-trivial kernel relations.
\par 
It has turned out that {\em Descent Theory} \cite{Groth1959} is a good tool for quantifying the interrelation of kernel pairs on a common domain. In Section \ref{section2} we describe descent data and point out its two main facets: On the one hand it describes algebraic structures, on the other hand it codes lifted equivalence relations in pullback squares. 
\par 
In Section \ref{section4} we introduce precise notions of {\em reachability} of pullback spans and of {\em coherence} of a pair of algebraic structures. Algebraic structures are coherent if they are reducts of a uniquely determined larger algebraic structure. Thus coherence is a local property in the sense that it can be cheked by investigating the basic material only, whereas reachability is a global property which is hardly checkable. In the main contribution of this paper (Proposition \ref{prop3}) we prove reachability to be equivalent to coherence, if the specification square is a pushout. This provides an answer to Question 1. 
\par 
This answer is formulated in a practical way in Theorem \ref{theorem1}. It also yields an answer to Question 2 in Theorem \ref{theorem3} which is a surprising analogon to the amalgamation lemma in Figure \ref{Fig0b}. Unfortunately, Theorem \ref{theorem3} is still unpractical in that we still have to investigate {\em all} pullback spans in order to decide the Van Kampen property. But we can show that a practical answer to Question 2 can be achieved if matters are restricted to sets and graphs (Section \ref{section5}, Theorem \ref{theorem2}).
\par
As related work we want to mention that \cite{LS06} prove topoi to be adhesive with similar methods (i.e.\ they use descent theory and some similar auxiliary results). Furthermore, \cite{VKAsBicolimits} show that being a Van Kampen square in a category $\C$ is equivalent to saying that its embedding into a certain span category over $\C$ is a pushout. In contrast to this generalization to higher level structures, we aim at an elementary characterization which can be checked within $\C$. 

\section{Descent Theory}\label{section2}
In this section, we work in a general topos $\C$. We will use the following notations: $\Ob{\C}$, $\Mor{\C}$ denote objects and arrows of any category $\C$, resp. ''$\EpiMapping{}{}{}$'' denotes epimorphisms. $x\in \C$ means $x\in \Ob{\C}$. The application of a functor $\FunctorF$ to an object or an arrow $x$ will be denoted without parenthesis: $\FunctorF x$. For an arrow $p$ of $\C$ we sometimes want pullbacks along $p$ to be uniquely determined. Thus we work with {\em chosen} pullbacks. The notation for the pullback functor $p^\ast$ is
\[\xymatrix{
{E\times_BA}\ar[d]_{\pi_1^\alpha:=p^\ast\alpha}\ar[r]^(.6){\pi_2^\alpha}			& 	{A}\ar[d]^{\alpha}	\\
{E}\ar[r]^{p}								                						&	{B}	\\
}
\]
 where $(\pi_1^\a, \pi_2^\a)$ is the chosen pullback of $(\alpha, p)$ (emphasized by decorating projections with $\alpha$). 
\par
\xymatrixrowsep{2pc} 
\xymatrixcolsep{2pc} 
In an adjoint situation $\_\,\dashv\,\_$, $\eta$ is the unit, $\e$ the co-unit. If $p:E\to B$ is any arrow in a category with pullbacks and $p_\ast:\CO{E}\to \CO{B}$ is the post-composing-functor, we have $p_\ast\dashv p^\ast$. The monad arising from this adjunction is $(\FunctorT^p, \eta^p, \mu^p)$, i.e., $\FunctorT^p := p^\ast\circ p_\ast:\CO{E}\to \CO{E}$, $\eta^p := \eta$, and $\mu^p := p^\ast\e_{p_\ast}$.
\par 
We intend to describe the categories $des(p)$ of {\em Descent Data}, where $p:E\to B$ is an arrow in $\C$. Grothendieck invented this theory in order to reason about structures in $\CO{B}$ (which may be difficult) by reasoning about monadic algebraic structures over $\CO{E}$, thus in a sense ''descending'' along $p$ \cite{Groth1959}. 
\par
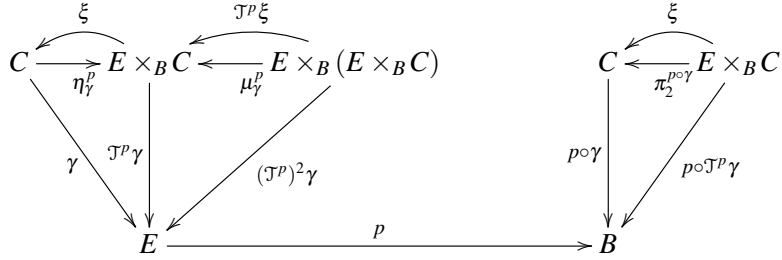
\begin{figure}[!htb]
\[
\xymatrix{
{C}\ar[ddr]_{\gamma}\ar[r]_{\eta^p_\gamma}		
								& {E\times_BC}\ar[dd]_{\FunctorT^p\gamma}\ar@/_1pc/[l]_{\xi}	
								& {E\times_B(E\times_BC)}\ar[ddl]^{(\FunctorT^p)^2\gamma}\ar@/_1pc/[l]_{\FunctorT^p\xi}\ar[l]^{\mu^p_\gamma} & {}		
								& C\ar[dd]_{p\circ \gamma}		
								& E\times_BC\ar[ddl]^{p\circ \FunctorT^p\gamma}\ar@/_1pc/[l]_{\xi}\ar[l]^{\pi_2^{p\circ \gamma}}			 \\
{}									& 		{}							& {}				& {}				& 		&		\\
{} 									& 		{E}\ar[rrr]^{p}					& {} 				& {}				& 	{B}	&		 \\
}
\]
\caption{Monadic Descent Data}\label{Fig1}
\end{figure}

We analyse the relationship between these algebraic structures and the category $pb(p)$ of all pullbacks along $p$ (to be defined precisely later on) such that it will facilitate our characterization of reachability in terms of descent data.
\begin{definition}[Descent Data]\label{def1} Let $\xymatrix{C\ar[r]^\gamma& E\ar[r]^p & B}$ be given and $(\FunctorT^p, \eta^p, \mu^p)$ be the monad on $\CO{E}$ arising from the adjunction $p_\ast \dashv p^\ast$. {\em Descent data} for $\gamma$ relative to $p$ is an arrow
\[\xi: \pi_1^{p\circ\gamma} = \FunctorT^p\gamma\to \gamma\]
of $\CO{E}$ with
\begin{equation}\label{eqn1}\xi\circ \eta^p_\gamma = id_C \mbox{ and } \xi \circ \FunctorT^p\xi = \xi \circ \mu^p_\gamma.
\end{equation}
\end{definition}
The situation is as in Figure \ref{Fig1}.
\par
Besides the $\CO{B}$-arrow $\pi_2^{p\circ\gamma}$, the right-hand side shows objects and the arrow $\xi$ after applying the left-adjoint $p_\ast$ only ($p_{\ast}$ is the identity on arrows of $\CO{E}$). Note that $\pi_2^{p\circ\gamma}$ establishes the co-unit of the adjunction $p_\ast\dashv p^\ast$. Thus
\begin{equation}\label{eqn2.7}\pi_2^{p\circ\gamma}\circ\eta^p_\gamma = id\mbox{ and }\mu^p_\gamma = p^\ast\pi_2^{p\circ\gamma}.
\end{equation}
Note that, for some $\gamma$ and $p$, an arrow $\xi$ as in Definition \ref{def1} must not exist and must not be unique. For future reference, we note that the $E\times_BC$-endomorphism $\o{\xi}:= \la\gamma\circ\pi_2^{p\circ\gamma}, \xi\ra$ can reconstruct $\xi$ via
\begin{equation}\label{eqn2.75}\xi = \pi_2^{p\circ\gamma}\circ \o{\xi}.\end{equation}
\cite{JT1994} gives a detailed investigation on that topic. It is also shown that
\begin{equation}\label{eqn2.76}\o{\xi} \circ \o{\xi} = id_{E\times_BC}.\end{equation}

\begin{definition}[Category of Descent Data]\label{def2} The category $des(p)$ has objects $(\gamma, \xi)$ with the properties of Definition \ref{def1} and arrows $h:(\gamma, \xi)\to(\gamma',\xi')$ the morphisms $h:\gamma\to \gamma'$ of $\CO{E}$ with $\xi'\circ \FunctorT^ph = h\circ \xi$.
\end{definition}
\begin{definition}[Category of Pullbacks]\label{def} For any $\Mapping{E}{p}{B}\in \Mor{\C}$ let $pb(p)$ denote the category with objects commutative diagrams of arbitrary pullbacks \footnote{
	... not only chosen pullbacks ...
} along $p$ together with morphism pairs $(m_1, m_2) \in Mor_{\C\downarrow{E}}\times Mor_{\C\downarrow{B}}$ such that the rear square in Figure \ref{Fig1,5} commutes. Note that by the decomposition property of pullbacks the rear square is a pullback, too.
\end{definition}
\xymatrixrowsep{1.5pc} 
\xymatrixcolsep{1.5pc} 
 \begin{figure}[H]
 \[
 \xymatrix{
 C\ar[rr]^q\ar[dd]_\gamma\ar[dr]^{m_1}	& 										& A\ar[dd]^<<<<<<{\a}\ar[dr]^{m_2}& 					 \\
 										& C'\ar[rr]^<<<<<<{q'}|{\hole}\ar[dl]^{\gamma'}	& 						& A'\ar[dl]^{\a'}	 \\
 E\ar[rr]_p								& 										& B								& 					 \\				
 }\]
 \caption{The category $pb(p)$}\label{Fig1,5}
 \end{figure}
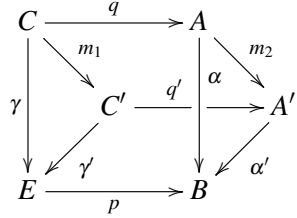
\xymatrixrowsep{2pc} 
\xymatrixcolsep{2pc} 

The monoidal conditions \myref{eqn1} (neutrality and associativity) imply that $des(p)$ is the {\em Eilenberg-Moore Category} associated with the monad $\FunctorT^p$. Thus, there is the comparison functor $\Phi^p: \CO{B}\to des(p)$ \cite{BaWe1990}. Obviously $pb(p)$ is equivalent to $\CO{B}$ via chosen pullbacks, such that we obtain a functor\footnote{
	To simplify matters, we still use the name $\Phi^p$ for this functor.
}
\[\Phi^p: pb(p)\to des(p).\]
In order to compute this functor, let us consider an arbitrary pullback $(q,\gamma)$ of a co-span $(p,\a)$ in $\C$, cf.\ Fig. \ref{Fig1,8}. Computing $\FunctorT^p\gamma$ using the chosen pullback $(p^\ast p_\ast\gamma = \FunctorT^p\gamma, \pi_2:=\pi_2^{p\circ\gamma})$ of $(p\circ\gamma ,p)$ yields a unique $\xi^\a:\FunctorT^p\gamma\to\gamma$ such that
\begin{equation}\label{eqn-canonical-descent1}
    q\circ\xi^\alpha = q\circ \pi_2.
\end{equation}
From \myref{eqn2.7}, \myref{eqn-canonical-descent1}, and the uniqueness of mediating morphisms for the original pullback one easily deduces
\[\xi^\a\circ \eta_\gamma^p = id_C.\]
Let $\o{\pi}_2:= \pi_2^{p\circ\FunctorT^p\gamma}$, then $\FunctorT^p\xi^\a:(\FunctorT^p)^2\gamma\to \FunctorT^p\gamma$ is unique with $\xi^\a\circ \o{\pi}_2 = \pi_2\circ \FunctorT^p\xi^\a$, such that a similar argumentation together with the second equation in \myref{eqn2.7} and \myref{eqn-canonical-descent1} yields
\[\xi^\a\circ \FunctorT^p\xi^\a = \xi^\a\circ \mu_\gamma^p.\]
Hence $\xi^\a$ fulfills \myref{eqn1}. Thus the original pullback is mapped to $(\gamma,\xi^\a)$, an object of $des(p)$. An investigation of the general construction of $\Phi^p$ \cite{BaWe1990} shows that our mapping reflects this construction where  
\begin{equation}\label{eqn2.79}
\Phi^p(m_1,m_2) = (m_1, \FunctorT^pm_1)
\end{equation}
on arrows. In the sequel, $(\gamma,\xi^\a)$ (or just $\xi^\a$ if $\gamma$ is fixed) will be called {\em canonical descent data} for the pullback of $\alpha$ along $p$.
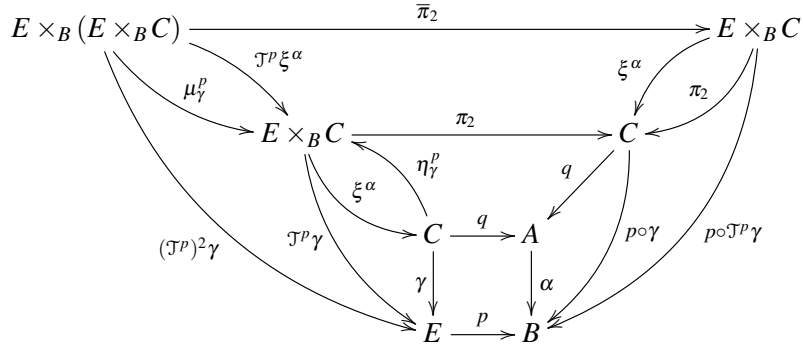
\begin{figure}[!htb]
\[\xymatrix{
{E\times_B(E\times_BC)}\ar@/_2pc/[dddrr]_{(\FunctorT^p)^2\gamma}\ar@/^1pc/[rd]^(.7){\FunctorT^p\xi^\a}\ar@/_1pc/[rd]^(.5){\mu^p_\gamma}\ar[rrrrr]^{\o{\pi}_2}
&&&&&{E\times_BC}\ar@/^2pc/[dddll]^(.5){p\circ\FunctorT^p\gamma}\ar@/_1pc/[dl]_(.7){\xi^\alpha}\ar@/^1pc/[dl]_(.5){\pi_2}
\\	
&    {E\times_BC}\ar@/_1pc/[ddr]_(.4){\FunctorT^p\gamma}\ar@/_1pc/[rd]^(.5){\xi^\alpha}\ar[rrr]^{\pi_2}
&&&  C\ar@/^1pc/[ddl]^(.4){p\circ \gamma}\ar[dl]_(.5){q}		\\
&&   C\ar[r]^q\ar[d]_{\gamma}\ar@/_1pc/[ul]_>>>>>>>{\eta_\gamma^p}
&    A\ar[d]^{\alpha}\\
&&   {E}\ar[r]^{p}				
&    B	&		 \\
}\]
\caption{Canonical Descent Data}\label{Fig1,8}
\end{figure}
\par
From Fig. \ref{Fig1}, we obtain $p\circ\gamma\circ \pi_2^{p\circ\gamma} = p\circ\gamma\circ\xi$ for each $(\gamma,\xi)\in \Ob{des(p)}$. Hence there is a functor $\Psi^p:des(p)\to \CO{B}$ which maps $(\gamma,\xi)$ to the unique arrow $\alpha$, which mediates $p\circ\gamma$ and a chosen coequalizer $c$ of $\pi_2^{p\circ\gamma}$ and $\xi$ (cf.\ also Fig. \ref{Fig3}). \cite{JT1994} shows that
\begin{itemize}\label{JT}
\item[i)] $\Psi^p$ is left-adjoint to the comparison functor $\Phi^p:\CO{B}\to des(p)$ with monic co-unit and
\item[ii)] if $p$ is an epimorphism, $\Phi^p$ becomes an equivalence of categories with pseudo-invers $\Psi^p$.
\end{itemize}
We use these facts to state the main result of this section. For this, we need some auxiliary considerations. The following statement is Lemma 20 in \cite{LS06}:
\par
\noindent
\begin{minipage}[t]{11cm}
\begin{lemma}\label{lemma1} Let $\C$ be a topos and a commutative diagram be given with an epimorphism as indicated. If $(1)+(2)$ and $(1)$ are pullbacks, then $(2)$ is a pullback, too.
\end{lemma}
\end{minipage}
\rule{1.0cm}{0cm}\hfill
\begin{minipage}[t]{3.9cm}
$\xymatrix{
	{\cdot}\ar[r]\ar[d]\ar@{}[dr]|{(1)}	& {\cdot}\ar[r]\ar[d]\ar@{}[dr]|{(2)}	& {\cdot}\ar[d]	\\
	{\cdot}\ar@{->>}[r]					& {\cdot}\ar[r]							& {\cdot}		\\
}$
\end{minipage}
\par\vspace{2ex}
\noindent
\begin{definition}[Equivalence Relation] \label{Def3} An {\em equivalence relation} on $A\in \Ob{\C}$ is a pair of arrows $a,b:U\to A$, such that $\MonoMapping{U}{\la a,b\ra}{A\times A}$ is a monomorphism, and which is
\begin{enumerate}
\item reflexive: $\exists r:A\to U: a\circ r = b\circ r = id$,
\item symmetric: $\exists s:U\to U: a\circ s = b, b\circ s = a$, and
\item transitive: If $(p:P\to U,q:P\to U)$ is the pullback of $(a,b)$ (especially $b\circ p = a\circ q$), there is $t:P\to U$, such that $a\circ t = a\circ p$ and $b\circ t = b\circ q$.
\end{enumerate}
\end{definition}
 
\begin{lemma}\label{lemmaXiIsEqRel}
$\Mapping{E\times_BC}{\la\xi,\pi_2^{p\circ\gamma}\ra}{C\times C}$ establishes an equivalence relation.
\end{lemma}
{\em Proof:} Because $\xi: \pi_1^{p\circ\gamma} = \FunctorT^p\gamma \to \gamma$, it is not difficult to see, that $\la\xi,\pi_2^{p\circ\gamma}\ra$ is monic. For reflexivity, let $r:=\eta^p_\gamma$ and use \myref{eqn1} and \myref{eqn2.7}. Symmetry follows with $s:=\o{\xi}$, \myref{eqn2.75}, and \myref{eqn2.76}. Transitivity can be established via $t:=\mu_\gamma^p$ (using the commuting top square in Fig. \ref{Fig1,8} and \myref{eqn1}). 
\qed
Note that this implies that $\la\xi,\pi_2^{p\circ\gamma}\ra$ is the kernel pair of its coequalizer, because in topoi, equivalence relations are effective (see \cite{J2002}, A 2.4.1.).
Consider now the above introduced coequalizer construction for $\Psi^p$.
\begin{figure}[H]
\[
\xymatrix{
{E\times_BC}\ar[d]_{\pi_2^{p\circ\gamma}}\ar[r]^{\xi}		&	{C}\ar[d]^c\ar[r]^{\gamma}					& 	{E}\ar[d]^p	\\
{C}\ar@{->>}[r]_c 											&	{H}\ar[r]_{\alpha}							&	{B}	\\
}
\]
\caption{Coequalizer construction}\label{Fig3}
\end{figure}
\begin{lemma}\label{lemma2} The right square in Figure \ref{Fig3} is a pullback. Hence, $\Psi^p: des(p) \to pb(p)$\footnote{
	Again not changing the name for this new functor.
}.
\end{lemma}
{\em Proof:} By Lemma \ref{lemmaXiIsEqRel} and because equivalence relations are the kernel pair of their coequalizer, the left square in Figure \ref{Fig3} is a pullback. Because $\gamma\circ \xi = \FunctorT^p\gamma$ and $\alpha\circ c = p\circ\gamma$ by definition of $\alpha$, the outer rectangle in Figure \ref{Fig3} is the pullback of $p\circ\gamma$ and $p$ as indicated in Figure \ref{Fig1}. Since $c$ is epic, the result follows from Lemma \ref{lemma1}.
\qed
\begin{proposition}[Correspondence of Pullbacks and Descent Data]\label{prop1}\rule{0mm}{0mm}
\begin{itemize}
\item[a)] For each choice of coequalizer in the construction of $\Psi^p$ the unit of the adjunction $\Psi^p\dashv \Phi^p:pb(p)\to des(p)$ is the identity.
\item[b)] If $p$ is an epimorphism, $\Phi^p: pb(p)\to des(p)$ becomes an equivalence of categories. Moreover, the coequalizer in the construction of $\Psi^p$ can be chosen such that the co-unit is identical.
\end{itemize}
\end{proposition}
{\em Proof:} We use the facts i. and ii. on page \pageref{JT}. For a) we use Lemma \ref{lemma2}: If $(\gamma,\xi)\in \Ob{des(p)}$, $\Phi^p\Psi^p(\gamma, \xi)$ is unique with \myref{eqn-canonical-descent1} (with $q$ replaced by $c$) by the above considerations on $\Phi^p$. But the coequalizer construction also yields $c\circ \xi = c\circ \pi_2^{p\circ\gamma}$, such that $(\gamma,\xi) = \Phi^p\Psi^p(\gamma, \xi)$, hence the unit is the identity. To prove b) consider an arbitrary pullback square $sqr$ as in Figure \ref{Fig1,8}. Pullbacks in topoi preserve epimorphisms (\cite{Goldblatt1984-t}, 5.3) thus, using the isomorphic co-unit, it is easy to show, that the diagram $q\circ\xi^\a = q\circ\pi_2^{p\circ\gamma}$ in the upper right corner of Figure \ref{Fig1,8} establishes a coequalizer situation. Hence for this choice of coequalizer, $\Psi^p\Phi^p sqr = sqr$, yielding an identical co-unit.\qed
\par 
For future reference, we want to illustrate these facts in the category $SET$. In the following proposition, the first part reformulates neutrality and associativity, whereas the nature of descent data as equivalence relation (on $C$) becomes evident from the second part. For a detailed explanation of this proposition, the reader is referred to the Appendix.
\begin{proposition}[Descent Data in SET]\label{prop2} Let $\C = SET$. 
\begin{enumerate}
\item \label{prop2_1}There is a bijective correspondence between objects $(\gamma,\xi)$ of $des(p)$ and families $(\xi_{e,e'})_{(e,e')\in ker(p)}:\gamma^{-1}e\to \gamma^{-1}e'$ of bijections which satisfy 
\[\xi_{e,e} = id_{\gamma^{-1}e}\quad
\mbox{and}
\quad\xi_{e,e''} = \xi_{e',e''}\circ \xi_{e,e'}.\]
for all $(e,e'), (e, e'') \in ker(p)$. 
\item \label{prop2_2} Let $c$ be the coequalizer of $\xi$ and $\pi_2^{p\circ\gamma}$. Then 
\[ker(c) = \{(x, \xi_{\gamma(x),\gamma(y)}(x))\mid x,y\in C,\,(\gamma(x), \gamma(y))\in ker(p)\}.\]
\end{enumerate}
\end{proposition}

\section{Coherence and Van Kampen Squares}\label{section4}
\begin{figure}[H]
\[
\xymatrix{
{} 						& 		{I}\ar[dd]|{\hole}^>>>>>{\gamma}\ar@{-->}[dr]^{s'?}\ar[ld]_{a'}\ar[rr]^{r'}& {} & {H}\ar[dd]^{\beta}\ar@{-->}[dl]^{\o{a}'?}\\
{J}\ar[dd]^{\tau}\ar@{-->}[rr]_<<<<<<<{\o{r}'?}&	{} 												& {K}\ar@{-->}[dd]^<<<<<<<<{\sigma\,?}	& {} 			\\
{} 						& 		{L}\ar[rr]|{\hole}^<<<<<<<<{r}\ar[ld]_{a}\ar[dr]|{s:=\overline{a}\circ r}	& {} 				& {R} \ar[ld]_{\overline{a}}\\
{A}\ar[rr]_{\overline{r}}	& 		{} 							& {S} 	& {}  \\	
}
\]
\caption{Reachability}\label{Fig0}
\end{figure}
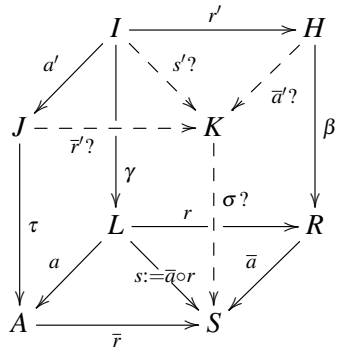

In this section we study the interplay of reachability of pullback spans and coherent coexistence of descent data in a general topos $\C$. After having defined these two concepts precisely, we state a local criterion for reachability and a global characterization of Van Kampen squares in terms of coherent algebraic structures. Let a commuting square as in the bottom of Figure \ref{Fig0} be given. 
\paragraph{Reachability:} Because $s = \o{r}\circ a$, we can decompose any diagonal pullback in $pb(s)$ into a left and a right part by calculating the right part via the chosen $\o{r}^\ast\sigma$. This calculation of the left part of the pullback of $pb(s)$ 
extends to a functor $\Delta^{\o{r}}:pb(s)\to pb(a)$. From $s = \o{a}\circ r$, we obtain $\Delta^{\o{a}}:pb(s)\to pb(r)$ in the same way. Then we define  
\[PB:=\la\Delta^{\o{r}}, \Delta^{\o{a}}\ra:pb(s) \to pb(a)\times_{\C\downarrow L}pb(r).\] 
where $pb(a)\times_{\C\downarrow{L}}pb(r)$ is the category of all pullback spans over $(a,r)$ together with morphism triples similar to the definition in Figure \ref{Fig1,5}. The name clash of this functor with the functor $PB$ in the introduction is deliberate: Both functors are equal up to an equivalence of categories, because $\CO{S}\cong pb(s)$.  
\begin{definition}[Reachability]\label{def4} A pullback span in $pb(a)\times_{\C\downarrow L}pb(r)$ is said to be reachable, if it is in the image of $PB$ up to a $pb(a)\times_{\C\downarrow L}pb(r)$-isomorphism.
\end{definition}
\paragraph{Coherence:} To investigate the counterpart of reachability on the instance level, we apply the methodology of Section \ref{section2} to the situation in Figure \ref{Fig0} in which the two back faces are pullbacks. Let $f:L\to B, g:B\to S$ be any two arrows in $\C$ and let $h:= g\circ f$. We consider the pullbacks $f^\ast(f\circ\gamma)$ and $h^\ast(h\circ\gamma)$ as in Figure \ref{Fig1} (with $C:=I$, $E:=L$, and $p:E\to B$ replaced by $f:L\to B$, $h:L\to S$, resp.). Let $\pi_2^{f\circ\gamma}, \pi_2^{h\circ\gamma}$ be the ''second projections'' in these pullbacks, resp.
\par 
For any $\gamma\in \CO{L}$ we have $h\circ\gamma\circ \pi_2^{f\circ\gamma} = h\circ\FunctorT^f\gamma$, thus there is a {\em unique} $u^g_\gamma : L\times_BI\to L\times_SI$ with
\begin{equation}\label{eqn10}\pi_2^{h\circ\gamma}\circ u^g_\gamma = \pi_2^{f\circ\gamma}\mbox{ and }\FunctorT^h\gamma\circ u^g_\gamma = \FunctorT^f\gamma
\end{equation}
cf.\ Figure \ref{Fig8}.
\begin{figure}[H]
\[
\xymatrix{
{L\times_BI}\ar@/_1pc/[ddr]_{\FunctorT^f\gamma}\ar@/^1pc/[drr]^{\pi_2^{f\circ\gamma}}\ar@{ (-->}[dr]_{u^g_\gamma}	& 	&						 \\
											& {L\times_SI}\ar[r]_{\pi_2^{h\circ\gamma}}\ar[d]_{\FunctorT^h\gamma}		& {I}\ar[d]^{h\circ\gamma}\\
											& L\ar[r]_{h}													& {S}					 \\
}\]
\caption{Construction of Embedding}\label{Fig8}
\end{figure}
Note, that in $SET$, $\FunctorT^f\gamma$ and $\FunctorT^h\gamma$ are first projections, which actually makes $u^g$ invariant under projections: Indeed $L\times_BI = \{(l,i)\,|\, f(\gamma(i)) = f(l)\}\subseteq \{(l,i)\,|\, h(\gamma(i)) = h(l)\} = L\times_SI$ where the embedding is $u^g$. This justifies the use of the hooked arrow in Figure \ref{Fig8}.
\par
In this way, we obtain $5$ embeddings for the original pushout situation: 
\[u^{\o{r}}_\gamma : L\times_AI\to L\times_SI, \; u^{\o{a}}_\gamma : L\times_RI\to L\times_SI\]
(using $s = \o{a}\circ r = \o{r}\circ a$ instead of $h= g\circ f$) as well as
\[u^r_\gamma : L\times_LI\to L\times_RI, \; u^a_\gamma : L\times_LI\to L\times_AI, \; u^s_\gamma : L\times_LI\to L\times_SI\]
(using $r = r\circ id_L$, $a = a\circ id_L$, and $s = s\circ id_L$) with corresponding projection compatibility and uniqueness as in \myref{eqn10}. The uniqueness property easily yields compositionality: 
\begin{equation}\label{eqn4.56}
\forall \gamma\in\CO{L}: u^s_\gamma = u^a_\gamma\circ u^{\o{r}}_\gamma = u^r_\gamma\circ u^{\o{a}}_\gamma. 
\end{equation}  
It can easily be shown that $u^g_\gamma$ are monomorphisms, but we can do better (see the Appendix for a proof): 
\begin{lemma}\label{lemmaMonadMorphism} Let $\xymatrix{L\ar[r]^f	& B \ar[r]^g & S}$ be given with $h:=g\circ f$. $u^g:\FunctorT^f\Rightarrow \FunctorT^h$ is a monad monomorphism.
\end{lemma}
\begin{lemma}\label{lemmaDescentForget}
Let $f,g,h$ be as in Lemma \ref{lemmaMonadMorphism}. There is a full and faithful functor $U^g:des(h)\to des(f)$ for which  
\[U^g(\gamma, \xi) = (\gamma, \xi\circ u^g_\gamma).\]
\end{lemma}
{\em Proof:} Since $des(p)$ is the category of Eilenberg-Moore-Algebras associated with $\FunctorT^p$, the result follows from Lemma \ref{lemmaMonadMorphism} and the proof of a theorem of Barr and Wells (\cite{TTT2005}, Theorem 6.3 in Chapter 3). \qed
\par 
Let us fix the rear pullback span $PBS$ in Figure \ref{Fig0}. Since $\gamma$ is fixed, considered objects of $des(r)$, $des(a)$, and $des(s)$ will always have codomain $\gamma$, hence $\xi^\beta$ and $\xi^\tau$ are appropriate abbreviations for the two canonical descent datas (cf.\ Section \ref{section2}) arising from the two pullbacks. 
\begin{definition}[Coherence]\label{def5}
$\xi^\tau$ and $\xi^\beta$ are called {\em coherent}, if there is $(\gamma,\xi)\in des(s)$, such that
\begin{equation}\label{eqn11}
	\la U^{\o{r}}, U^{\o{a}}\ra(\gamma, \xi) = (\xi^\tau, \xi^\beta)
\end{equation}
We call any $(\gamma,\xi)\in des(s)$ with this property a {\em coherence witness} (for $\xi^\tau$ and $\xi^\beta$).   
\end{definition}
Thus two algebraic structures are coherent, if there is an algebraic structure over $\gamma$ relative to $s$ which effectively approximates them. We are ready to state the main technical result of this section:
\begin{proposition}[Reachability vs. Coherence]\label{prop3}
Let in a topos $\C$ a diagram be given as in Figure \ref{Fig0} where the bottom square is commutative and the rear faces form a pullback span. Let $\xi^\tau$ and $\xi^\beta$ be the above introduced canonical descent datas.
\begin{itemize}
\item[a)] If the span is reachable, $\xi^\tau$ and $\xi^\beta$ are coherent.
\item[b)] If the bottom square is a pushout and $\xi^\tau$ and $\xi^\beta$ are coherent, then the span is reachable.
\item[c)] Under the prerequisites of b), the coherence witness is unique.   
\end{itemize}
\end{proposition}
{\em Proof:} To simplify matters we write $u$ instead of $u_\gamma$. To show a), let $sqr_{diag}\in pb(s)$ (the pullback $(\gamma, s')$ of $(\sigma, s)$) with $PB(sqr_{diag})$ being the rear pullback span (this is Figure \ref{Fig0} without question marks)\footnote{
	If $PB(sqr_{diag})$ yields the rear pullback span not exactly but only up to isomorphism, we can exchange the instances over $A$ and $R$ by their compositions with the isomorphisms, such that there is a complete cube with 4 pullbacks having the original pullback span as rear faces. It is no problem that front and right pullbacks are no longer chosen.  
}. We show coherence with $\xi:=(\gamma,\xi):=\Phi^s sqr_{diag}$. By \myref{eqn-canonical-descent1} $\xi^\tau$ is unique with $a'\circ\xi^\tau = a'\circ\pi_2^{a\circ\gamma}$, such that for the first projection in \myref{eqn11} it suffices to show validity of this equation with $\xi^\tau$ replaced by $\xi\circ u^{\o{r}}$. The argumentation for the second projection is then similar. We have
\begin{align*}
\tau\circ a'\circ\xi\circ u^{\o{r}}	&= a\circ\gamma\circ\xi\circ u^{\o{r}}		&& \mbox{Left rear pullback in Figure \ref{Fig0}}			 \\
				                        &= a\circ\FunctorT^s\gamma\circ u^{\o{r}}	&& \mbox{Since }\xi:\FunctorT^s\gamma \to \gamma		 \\
				                        &= a\circ\FunctorT^a\gamma					&& \mbox{By \myref{eqn10}}							 \\
				                        &= a\circ\gamma\circ\xi^\tau				&& \mbox{Since }\xi^\tau:\FunctorT^a\gamma \to \gamma	\\
				                        &= \tau\circ a'\circ\pi_2^{a\circ\gamma}	&& \mbox{Left rear pullback and \myref{eqn-canonical-descent1} for }\xi^\tau\\
\end{align*}
and also $\o{r}'\circ a'\circ\xi\circ u^{\o{r}} = s'\circ\pi_2^{s\circ\gamma}\circ u^{\o{r}} = s'\circ\pi_2^{a\circ\gamma} = \o{r}'\circ a' \circ \pi_2^{a\circ\gamma}$ (by \myref{eqn-canonical-descent1} for $\xi$ and \myref{eqn10}). This implies the desired result, because in the front face pullback $\tau$ and $\o{r}'$ are jointly monic.
\par
To show b), asssume we already knew the result in the case $a$ and $r$ are both epimorphisms. We can then use epi-mono-factorizations $r = r_m\circ r_e$ and $a = a_m\circ a_e$ (which exist in topoi) to decompose both back face pullbacks into two pullbacks resp. It can then be verified that the bottom face can be decomposed into 4 pushouts along these epi-mono-factorizations. Because $r_e$ and $a_e$ are both epic and $\xi^\tau$ and $\xi^\beta$ are also canonical descent datas of $r_e$ and $a_e$ (this follows from fact i.\ on page \pageref{JT}), the inner pullback span is reachable. Since topoi are adhesive \cite{LS06}, this reachability can be continued along the other pairs of bottom arrows (of which either one or both are now monic) by constructing top face pushouts.
\par
Thus it suffices to assume that $r$ and $a$ are epic. Then $r'$ in Figure \ref{Fig0} is the appropriate coequalizer of $\xi^\beta$ and $\pi_2^{r\circ \gamma}$ by Proposition \ref{prop1}, b). Let $s':I\to K$ be the coequalizer of $\pi_2^{s\circ \gamma}$ and the coherence witness $\xi$ with $sqr_{diag}:=\Psi^s(\gamma,\xi)$ the resulting diagonal pullback by Lemma \ref{lemma2}. By coherence and \myref{eqn10}
\[s'\circ \xi^\beta = s'\circ \pi_2^{r\circ\gamma}\]
yielding a unique mediator $\o{a}':H\to K$ for the coequalizer $r'$, i.e.
\begin{equation}\label{eqn3.1}\o{a}' \circ r' = s'.\end{equation}
Let $\sigma\in\CO{S}$ be part of $sqr_{diag}$ as indicated in Figure \ref{Fig0}. Then by construction and \myref{eqn3.1} $\o{a}\circ\beta\circ r' =\o{a}\circ r\circ\gamma = s\circ \gamma = \sigma\circ s' = \sigma\circ\o{a}' \circ r'$, hence we obtain a commutative square as right face of the cube in Figure \ref{Fig0} (the coequalizer $r'$ is an epimorphism) which is also a pullback by Lemma \ref{lemma1}, i.e. $\o{a}^\ast\sigma \cong \beta$. Analogously one shows $\o{r}^\ast\sigma \cong \tau$. 
\par 
To show c) assume that there are two coherence witnesses $(\gamma_1, \xi_1), (\gamma_2, \xi_2)$. Clearly $\gamma:=\gamma_1 = \gamma_2$ by Lemma \ref{lemmaDescentForget}, such that it remains to show $\xi_1 = \xi_2$. By b), $\xi_1$ and $\xi_2$ yield two cubes each of which possess $4$ pullbacks as side faces. They possess the same arrows except $\o{a}', \o{r}'$, and $\sigma$. But the two variants of the arrows $\o{a}', \o{r}'$ both form a top pushout of $a', r'$ because, in topoi, pullbacks preserve colimits. Hence there is an isomorphism $i$ which can be shown to mediate between the two variants of $\sigma$. 
\par 
Consequently, we have two diagonal pullbacks $sqr^1_{diag} = \Psi^s(\gamma, \xi_1)$ and $sqr^2_{diag} = \Psi^s(\gamma, \xi_2)$ (see part b)) for which by \myref{eqn2.79} and Proposition \ref{prop1} a) 
\[\Phi^s(\Mapping{sqr^1_{diag}}{(id,i)}{sqr^2_{diag}})= \Mapping{(\gamma, \xi_1)}{(id,\FunctorT^sid)}{(\gamma, \xi_2)}\]
which yields $\xi_1 = \xi_2$.  
\qed
By the remark after Lemma \ref{lemmaXiIsEqRel} any descent data $(\gamma, \xi)\in des(p)$ yields the kernel pair $ker(q):=(\xi, \pi_2)$ of the top arrow $q$ of $\Psi^p(\gamma, \xi)$, see Figure \ref{Fig1,8}. In the category $Eq(C)$ of equivalence relations on $C\in \Ob{\C}$ (i.e. the full subcategory of $\CO{(C\times C)}$ of arrows with the properties of Definition \ref{Def3}), we call an object $e$ an upper bound of $e_1$ and $e_2$, if there are $Eq(C)$-arrows (necessarily monos) $v_1$ and $v_2$ with 
\begin{equation}\label{eqn3.65}e\circ v_1 = e_1 \mbox{ and } e\circ v_2 = e_2.\end{equation}
It is well-known \cite{TTT2005} that the least upper bound ($lub$) of two equivalence relations $ker(a'):X\to C\times C, ker(r'): Y\to C\times C$ can be constructed by extracting the mono part $m$ of $[ker(a'), ker(r')]: X + Y \to C\times C$ followed by constructing the kernel pair of the coequalizer of $m$. Let $\pi_2'$ be the second projection in the pullback associated with the monad $\FunctorT^s$.  
\begin{lemma}\label{lemmaCoherenceAndLUB} Let the bottom square in Figure \ref{Fig0} be a pushout. $\xi^\tau$ and $\xi^\beta$ are coherent if and only if there is $(\gamma, \xi)\in des(s)$ with  
\[lub(ker(a'), ker(r')) \cong (\xi, \pi_2')\]
\end{lemma}   
{\em Proof:} If $\xi^\tau$ and $\xi^\beta$ are coherent, then the coherence witness $\xi$ from Proposition \ref{prop3} b) and c) was used to complete the pullback cube. Since, in topoi, the top face becomes a pushout, and $(\xi,\pi_2')$ is the kernel pair of the top diagonal, using the universal property of pushouts, it can easily be shown that $(\xi,\pi_2') \cong lub(ker(a'), ker(r'))$. 
\par
The opposite direction follows directly from the uniqueness properties \myref{eqn10} of the monad morphisms $u^{\o{a}}$ and $u^{\o{r}}$: Any mediating monomorphisms $v_1, v_2$ as in \myref{eqn3.65} in the least upper bound constellation must coincide with $u^{\o{a}}$, $u^{\o{r}}$, resp.
\qed
\begin{theorem}[Answer to Question 1]\label{theorem1} Let $\C$ be a topos and a pullback span be given as in the rear of Figure \ref{Fig0} with the bottom square a pushout. The span is reachable if and only if $lub(ker(a'),ker(r')) \cong (\xi, \pi_2')$ for some $(\gamma, \xi)\in des(s)$. 
\end{theorem}
{\em Proof:} This follows from Proposition \ref{prop3} and Lemma \ref{lemmaCoherenceAndLUB}. \qed
Thus there is an algorithm to check reachability: Given a rear pullback span $PBS$ with top arrows $a', r'$
\begin{enumerate}
\item Compute $e:=lub(ker(a'), ker(r'))$.
\item Check the monadicity requirements \myref{eqn1} of $e$ relative to $s$ by interpreting it as a pair $(\xi, \pi_2')$. 
\item $PBS$ is reachable if and only if $e$ meets the requirements.  
\end{enumerate} 
In the next section we will recall the introductory examples from Section \ref{section1,5} such that these theoretical results become more evident. 
\par 
We conclude this section with a global statement on Van Kampen squares in the spirit of Figure \ref{Fig0b}. We still assume a square as the bottom in Figure \ref{Fig0} to be given. In the following, the category $des(id_L)$ is integrated. It represents the ''common part'' of the forgetful functors $U^a$ and $U^r$, namely the carrier $\gamma$ represented by certain isomorphisms from the ''graph'' $L\times_LI$ of $\gamma$ to $I$. 
\begin{theorem}[Fibred Version of Amalgamation Lemma]\label{theorem3} Let $\C$ be a topos. In Figure \ref{Fig0c}, the pushout (1) is a Van Kampen square if and only if (2) is a pullback in $CAT$.
\end{theorem}
\begin{figure}[!htb]
\[\xymatrix{
	L\ar[rr]^r\ar[dd]_a	&&	R	\ar[dd]_{\o{a}}		&&&	des(id_L)\ar@{}[ddrr]|{(2)}&& 	des(r)\ar[ll]_{U^r}													\\
											&&											&&&													&&																							\\
	A\ar[rr]^{\o{r}}		&&	S\ar@{}[uull]|{(1)} &&&	des(a)\ar[uu]^{U^a}			&& des(s)\ar[uu]^{U^{\o{a}}}\ar[ll]_{U^{\o{r}}} \\
}\] 
\caption{Amalgamation Lemma (Fibred setting)}\label{Fig0c}
\end{figure}
{\em Proof:}
\par  
''$\Rightarrow$'': By \myref{eqn4.56} and Lemma \ref{lemmaDescentForget} (2) commutes. By assumption, $PB$ is an equivalence of categories, i.e.\ each rear pullback span is reachable. By Proposition \ref{prop3} a) each pair $(\xi^\tau, \xi^\beta)$ is coherent and by Proposition \ref{prop3} c) the coherence witness is unique. Standard arguments together with the fact that $U^{g}$ are full and faithful functors (Lemma \ref{lemmaDescentForget}) yield the pullback property.  
\par 
''$\Leftarrow$'': The pullback property immediately yields coherence for each pair $((\gamma, \xi^\tau), (\gamma, \xi^\beta)) \in des(a)\times des(r)$. Because (1) is a pushout, Proposition \ref{prop3} b) implies reachability of each rear pullback span, thus $PB$ is essentially surjective, which is sufficient for (1) to have the Van Kampen property \cite{Sobo2004}. \qed
  
\section{Coherence and Van Kampen Squares in SET and GRAPH}\label{section5}
This section illustrates the use of Theorem \ref{theorem1} and develops a simply checkable characterization of Van Kampen squares in $SET$ and $GRAPH$ based on Theorem \ref{theorem3}. As mentioned before, in $SET$, the $u$'s are natural embeddings. Hence coherence (cf.\ Definition \ref{def5}) yields the existence of descent data $\xi$ relative to $s=\o{a}\circ r = \o{r}\circ a$ with
\begin{equation}\label{eqn11.5}\forall (x,x')\in ker(r): \xi_{x, x'} = \xi^\beta_{x, x'}\mbox{ and }\forall (y,y')\in ker(a): \xi_{y, y'} = \xi^\tau_{y, y'}\end{equation}
where all mappings are understood as the components of the families of bijections from Proposition \ref{prop2}.
\par
We can now observe Theorem \ref{theorem1} at work: Recall the situation in Figure \ref{FigExample1}. By Proposition \ref{prop2}, \ref{prop2_2} the canonical descent data $\xi^\beta$ and $\xi^\tau$ map along the dashed and dotted lines, resp. E.g. $\xi^\beta_{x,y}(1\hspace{-0.2em}:\hspace{-0.2em}x) = 2\hspace{-0.2em}:\hspace{-0.2em}y, \xi^\beta_{x,y}(2\hspace{-0.2em}:\hspace{-0.2em}x) = 1\hspace{-0.2em}:\hspace{-0.2em}y$. Reachability means that the least upper bound of the kernels of $a'$ and $r'$ yield a monadic structure $\xi$ relative to $s$. By \myref{eqn11.5} and hypothetical associativity (cf.\ Proposition \ref{prop2}) of $\xi$ the bijection $\xi_{x,y}$ must be equal to $\xi^\tau_{w,y}\circ \xi^\beta_{z,w}\circ \xi^\tau_{x,z}$ on the fibre over $x$. But this {\em must} then coincide with $\xi^\beta_{x,y}$, which is not the case in Figure \ref{FigExample1}. 
\par
Obviously, the kernels of $r$ and $a$ are intertwined through the cycle $(x,z),(z,w),(w,y),(y,x)\in ker(s)$ and are thus not enough separated. The following definition makes this more precise:
\begin{definition}[Separated Kernels]\label{def5.9} Let $\C = SET$ and $a$ and $r$ be given as in Figure \ref{Fig0}. A sequence $(x_i)_{i\in\{0, 1, \ldots, 2k+1\}}$ of elements in $L$ is called a {\em domain cycle} (of $a$ and $r$), if $k\in\N$ and the following conditions hold:
\begin{enumerate}
\item\label{cyc1} $\forall j\in \{0,1,\ldots, 2k+1\}: x_j\not=x_{j+1}$
\item\label{cyc2} $\forall i\in\{0,\ldots, k\}: (x_{2i},x_{2i+1})\in ker(a)$
\item\label{cyc3} $\forall i\in\{0,\ldots, k\}: (x_{2i+1},x_{2i+2})\in ker(r)$
\end{enumerate}
where the sums are understood modulo $2k+2$ (i.e. $x_{2k+2} = x_0$). We call $2k+2$ the {\em length} of the domain cycle. Moreover, a domain cycle is {\em proper} if we have for all $i,j\in \{0,1,\ldots, 2k+1\}$ that $x_i\not= x_j$ if $i\not= j$.
\par
The pair $a$ and $r$ is said to have {\em separated kernels}, if it has no domain cycle.
\end{definition}
{\em Remark 1}: It is easy to see that each domain cycle $c$ possesses a proper subcycle, i.e.\ a proper cycle with smaller or equal length than the length of $c$ and whose elements are a subset of the elements of $c$.\\ 
{\em Remark 2}:
''Having separated kernels'' is only sufficient but not necessary for ''being jointly monic''. Indeed, being not jointly monic induces a domain cycle of length $2$. But longer domain cycles occur for jointly monic $a$ and $r$ (see Figure \ref{FigExample1}). 
\par 
Domain cycles are connected to coherence as follows:
\begin{proposition}\label{prop4} Let $\C = SET$ and a commutative square be given like the bottom square in Figure \ref{Fig0} and let the two rear faces be pullbacks with canonical descent data $\xi^\tau$ and $\xi^\beta$, resp. $\xi^\tau$ and $\xi^\beta$ are coherent iff for all domain cycles $(x_i)_{i\in\{0, 1, \ldots, 2k+1\}}$ of $a$ and $r$ we have
\begin{equation}\label{eqn-condition-coherence} \xi^\beta_{x_{2k+1},x_{0}}\circ\xi^\tau_{x_{2k},x_{2k+1}}\circ \cdots  \circ \xi^\tau_{x_2,x_3}\circ\xi^\beta_{x_1,x_2}\circ \xi^\tau_{x_0,x_1} = id_{\gamma^{-1}x_0}
\end{equation}
\end{proposition}
The statement is illustrated in Example \ref{Example2}, where coherence is now achieved by harmonizing the equivalences of $a'$ and $r'$ in the two copies of $L$ that make up the domain of $\gamma$. Alternatively, we can use Theorem \ref{theorem1} to check reachability: The least upper bound yields a descent data for $\gamma$ relative to $s$, because it is evident that neutrality and associativity are not destroyed. In order not to interrupt the flow of arguments, the proof of Proposition \ref{prop4} is contained in the Appendix. 
\par 
The next proposition illustrates how domain cycles are connected to reachability. This time we include the proof, because it demonstrates the use of descent data. 
\begin{proposition}\label{prop5} Let $\C = SET$ and a commutative square be given like the bottom square in Figure \ref{Fig0}. If all pullback spans in the rear are reachable, $a$ and $r$ have separated kernels.
\end{proposition}
{\em Proof:} Assume to the contrary that $a$ and $r$ possess a domain cycle $(x_i)_{i\in\{0, 1, \ldots, 2k+1\}}$ for some $k\in\N$. By the first remark after Definition \ref{def5.9}, we can assume that this cycle is proper. Let $\Omega = \{0,1\}$ and $\gamma:= \pi_2:\Omega\times L \to L$ be the ordinary second projection. We construct descent data $\xi^a$ for $\gamma$ relative to $a$ and $\xi^r$ for $\gamma$ relative to $r$: Because the fibre of $\gamma$ over $x$ is $\{(0,x), (1,x)\}$ we can define $\xi^r_{x,x'}(b,x):= (b,x')$ for all $(x,x')\in ker(r)$ and $b\in\{0,1\}$. It is obvious that this yields neutrality and associativity of Proposition \ref{prop2}.
\par
Consider now the equivalence class $E_0 = \{x\in L\mid a(x) = a(x_0)\}$ of $ker(a)$, where $x_0$ is the begin of the cycle. 
The domain cycle has at least length $2$, hence we have $x_1\not=x_0$, $x_1\in E_0$ in the cycle.
For any $x\in E_0$ we define a bijection $\xi^a_{x_0,x}:\{0,1\}\times\{x_0\}\to\{0,1\}\times\{x\}$ by
\[\xi^a_{x_0,x}(b,x_0):=\left\{
	\begin{array}{r@{\mbox{ if }}l}
		(b,x)	& 	x\not=x_1\\
		(1-b,x)	&	x=x_1
	\end{array}
\right.\]
Further we set 
\[\xi^a_{x,x'}:=\xi^a_{x_0,x'}\circ(\xi^a_{x_0,x})^{-1}\quad \mbox{for all} \quad x,x'\in E_0,\; x\not=x_0.
\]
 Neutrality and associativity are straightforwardly ensured by these definitons. For $(x,x')\in\ker(a) - E_0^2$ we define $\xi^a_{x,x'}$ in the same way as $\xi^r$.
\par
By Proposition \ref{prop1} a), $\xi^\beta:=\xi^r$ and $\xi^\tau:=\xi^a$ are canonical descent datas of the pullbacks $\Psi^r(\gamma, \xi^r)$ and $\Psi^a(\gamma, \xi^a)$, resp, such that for the resulting pullback span we obtain:
\[(\xi^\beta_{x_{2k+1},x_{0}}\circ\xi^\tau_{x_{2k},x_{2k+1}}\circ \cdots  \circ \xi^\tau_{x_2,x_3}\circ\xi^\beta_{x_1,x_2}\circ \xi^\tau_{x_0,x_1})(0,x_0) = (1,x_0)\]
because, in this chain, $\xi^\beta$ always preserves the first projection and $\xi^\tau_{x,x'}$ interchanges it only if $x = x_0$ and $x' = x_1$ since the cycle is proper. Thus, by Proposition \ref{prop4}, $\xi^\tau$ and $\xi^\beta$ are not coherent, hence, by Proposition \ref{prop3}, the pullback span is not reachable contradicting the assumption.\qed
The following theorem is the main result of this section (cf.\ \cite{L-10}):
\begin{theorem}\label{theorem2} 
Let $\C = SET$ or $\C = GRAPH$. A pushout diagram as the bottom square in Figure \ref{Fig0} is a Van Kampen square if and only if $a$ and $r$ have separated kernels.
\end{theorem}
{\em Proof:} ''$\Rightarrow$'' follows from Theorem \ref{theorem3} and Propositions \ref{prop3} and \ref{prop5}, ''$\Leftarrow$'' follows from Proposition \ref{prop4} and Theorem \ref{theorem3}. It is shown in \cite{L-10} that the argumentation easily carries over to graphs once the result has been proven for $SET$. \qed
Recall Example \ref{Example3}, where we can now easily derive from Theorem \ref{theorem2} that each pullback span is reachable, i.e.\ each amalgamation of instances is successful. 
\section{Outlook}\label{section6}

The paper presents first outcomes of a more comprehensive collaborative project based on \cite{WD2008_a,L-10, WD2007} and addressing "compositional fibred semantics in topoi". There are several topics for future research: First we have to address persistency requirements and extension lemmas for fibred semantics. Moreover, we are looking for a categorical generalization of Proposition \ref{prop4} which would give rise, due to Proposition \ref{prop3}, to a kind of "conditional compositionality". 
\par 
An interesting open question, in this context, is how to characterize domain cycles on a pure categorical level. This should yield an elementary characterization of Van Kampen squares in more general categories in the spirit of Theorem \ref{theorem2}.

\section{Appendix}\label{section7}
\paragraph{Descent Data in $SET$:}
Here, we give details about the different view on descent data from {\bf Proposition \ref{prop2}} in $SET$. First, we remind that pullbacks, in general, can be described as products in slice categories. For the situation in Definition \ref{def1} this means that the diagonal $p\circ\FunctorT^p\gamma=p\circ\gamma\circ\pi_2^{p\circ\gamma}:E\times_BC\to B$ forms the product $p\times (p\circ\gamma)$ in $\CO{B}$ with projections $\FunctorT^p\gamma:p\times (p\circ\gamma)\to p$ and $\pi_2^{p\circ\gamma}:p\times (p\circ\gamma)\to p\circ\gamma$. Second, any $\xi:E\times_BC\to C$ in $\C$ which is an arrow $\xi:\FunctorT^p\gamma\to\gamma$ in $\CO{E}$ establishes also an arrow $\xi:p\times (p\circ\gamma)\to p\circ\gamma$ in $\CO{B}$. $\CO{B}$, however, is also a topos by the fundamental theorem of Freyd \cite{Freyd:72} and thus, espcially cartesian closed. In $\C=SET$, finally, any $\xi \in Hom(p\times (p\circ\gamma),p\circ\gamma) \cong Hom(p,(p\circ\gamma)^{p\circ\gamma})$ can be interpreted as a map that assigns to any element $e\in E$ an endomap $\xi(e,.)$ of the fibre of $p\circ\gamma$ over $p(e)$ (cf.\ \cite{Goldblatt1984-t}, Chapter 4).
\par
For our purposes, an appropriate representation of these maps for descent data will be in terms of the kernel of $p$: The fibre of $p\circ\gamma$ over $p(e)$ is the pre-image of the equivalence class $[e]_{ker(p)}$ w.r.t.\ $\gamma$. Let $\xi_{e,e'}$ be the restriction of the map $\xi(e',.)$ to $\gamma^{-1}e$ whenever $(e,e')\in ker(p)$. If $c\in \gamma^{-1}e$ we obtain $\gamma(\xi(e',c)) = e'$ from Definition \ref{def1}, hence the codomain of $\xi_{e,e'}$ is $\gamma^{-1}e'$ and $\xi$ represents a family
\begin{equation}\label{eqn2.79a}(\xi_{e,e'}:\gamma^{-1}e\to \gamma^{-1}e')_{(e,e')\in ker(p)}\end{equation}
which fulfills
\begin{equation}\label{eqn2.79b}\xi(e',c) = \xi_{e,e'}(c)\mbox{ for }\gamma(c) = e.\end{equation}
Let us now investigate the influence of neutrality and associativity \myref{eqn1} to this family. A canonical choice of pullbacks in $SET$ yields
\begin{eqnarray*}
  E\times_BC            &=& \{(e,c)\in E\times C\mid p(e) = p(\gamma(c))\}, \\
  E\times_B(E\times_BC) &=& \{(e,(e',c))\in E\times(E\times C)\mid p(e) = p(e') = p(\gamma(c))\},
\end{eqnarray*}
and
\begin{equation}\label{eqn2.8}\eta^p_\gamma(c) = (\gamma(c), c),\, \mu^p_\gamma(e'',(e',c)) = (e'',c), \, \FunctorT^p\xi(e'',(e',c))=(e'',\xi(e',c)).
\end{equation}
Thus for all $(e,e'), (e',e'')\in ker(p)$ and $c\in\gamma^{-1}e$, \myref{eqn1} and the first equation in \myref{eqn2.8} yield
\[\xi_{e,e}(c) = c,\]
whereas the second equation in \myref{eqn1} (applied to a triple $(e'', (e',c))$) and the second and third equation of \myref{eqn2.8} imply
\[\xi_{e',e''}(\xi_{e,e'}(c))=\xi_{e,e''}(c).\]
By choosing $e'' = e$, these two equations force each $\xi_{e,e'}$ to be bijective.
\par
By reversing the whole argumentation, we can also show that any family as in \myref{eqn2.79a} which satisfies these two equations yields a descent data by defining $\xi$ as in \myref{eqn2.79b} for $e := \gamma(c)$. Altogether we obtain the statement in Proposition \ref{prop2}, \ref{prop2_1} which subsumes the monoidal nature of descent data in $SET$. Moreover, \ref{prop2_2} follows from effectiveness of equivalence relations and \myref{eqn2.79b}.

\paragraph{Proof of Lemma \ref{lemmaMonadMorphism}:}
For simplicity we write $u$ instead of $u^g$. There are several statements to prove: 
\begin{enumerate}
\item Each $u_\gamma$ is a monomorphism. 
\item $u:\FunctorT^f\Rightarrow \FunctorT^h$ is a natural transformation.
\item $u$ is compatible with units, i.e. $u\circ \eta^f = \eta^h$.
\item $u$ is compatible with co-units, i.e. $\mu^h\circ u^2 = u\circ \mu^f$ where $u^2$ is the horizontal composition of $u$ with itself.
\end{enumerate}
\par 
1. To show that $u_\gamma$ is monic for each $\gamma$, let $x,y:X\to L\times_BI$ with $u_\gamma\circ x = u_\gamma\circ y$ be given. By \myref{eqn10}, one computes $\FunctorT^f\gamma\circ x = \FunctorT^f\gamma\circ y$ and $\pi_2^{f\circ \gamma}\circ x = \pi_2^{f\circ \gamma}\circ y$. Because $\FunctorT^f\gamma$ and $\pi_2^{f\circ \gamma}$ are jointly monic (being a limit cone in a pullback square), we obtain $x = y$. In the sequel, the property of a pullback cone to be jointly monic will be used several times. We will do this without further reference.
\par
2. Let $\gamma, \hat{\gamma}\in \CO{L}$ and
\[\phi:\gamma\to \hat{\gamma}\]
be a $\CO{L}$-morphism. As before, $\pi_2$ and $\pi_2'$ denote the second projections in the pullbacks involving $\gamma$ and the monads $\FunctorT^f$ and $\FunctorT^h$, resp. $\hat{\pi}_2$ and $\hat{\pi}_2'$ denote the second projections involving $\hat{\gamma}$.
\par
Pulling back $\phi$ (as an arrow in $\CO{B}$ and as an arrow in $\CO{S}$) yields
\begin{equation}\label{eqn30a}
\phi\circ \pi_2 = \hat{\pi}_2\circ \FunctorT^f\phi
\end{equation}
and
\begin{equation}\label{eqn30b}
\phi\circ \pi'_2 = \hat{\pi}'_2\circ \FunctorT^h\phi
\end{equation}
Let now $d_1 = \FunctorT^h\phi\circ u_\gamma$ and $d_2 = u_{\hat{\gamma}}\circ \FunctorT^f\phi$, which are both arrows from $\FunctorT^f\gamma$ to $\FunctorT^h\hat{\gamma}$. $d_1 = d_2$ (and thus the desired result) follows from 
\begin{align*}
\FunctorT^h\hat{\gamma}\circ d_1 &= \FunctorT^h(\hat{\gamma}\circ\phi)\circ u_\gamma	&& \mbox{Definition of }d_1	\\
						&= \FunctorT^h\gamma\circ u_\gamma						&& \mbox{Because }\phi:\gamma\to \hat{\gamma}\\
						&= \FunctorT^f\gamma									&& \mbox{By \myref{eqn10}} \\
						&= \FunctorT^f\hat{\gamma}\circ \FunctorT^f\phi				&& \mbox{See two lines above} \\
						&= \FunctorT^h\hat{\gamma}\circ u_{\hat{\gamma}} \circ \FunctorT^f\phi && \mbox{By \myref{eqn10}} \\
						&= \FunctorT^h\hat{\gamma}\circ d_2 						&& \mbox{Definition of }d_2
\end{align*}
and
\begin{align*}
\hat{\pi}'_2\circ d_1  		&= \hat{\pi}'_2\circ \FunctorT^h\phi\circ u_\gamma			&& \mbox{Definition of }d_1	\\
						&= \phi\circ \pi'_2\circ u_\gamma						&& \mbox{By \myref{eqn30b}}	\\
						&= \phi\circ \pi_2									&& \mbox{By \myref{eqn10}}	\\
						&= \hat{\pi}_2\circ \FunctorT^f\phi						&& \mbox{By \myref{eqn30a}}	\\
						&= \hat{\pi}'_2\circ u_{\hat{\gamma}}\circ\FunctorT^f\phi		&& \mbox{By \myref{eqn10}}	\\
						&= \hat{\pi}'_2\circ d_2								&& \mbox{Definition of }d_2.	\\
\end{align*}
In the sequel we denote projections with $\pi_2, \o{\pi}_2$ in pullbacks along $f$ and with $\pi'_2, \o{\pi}'_2$ in pullbacks along $h$.
\par
3. Compatibility with the units follows from $\pi_2'\circ u_\gamma \circ \eta^f_\gamma = \pi_2\circ \eta^f_\gamma = id = \pi_2'\circ \eta^h_\gamma$ (apply \myref{eqn10} and \myref{eqn2.7} twice) and $\FunctorT^h\gamma\circ u_\gamma\circ \eta^f_\gamma = \FunctorT^f\gamma\circ\eta^f_\gamma = \gamma = \FunctorT^h\gamma\circ\eta^h_\gamma$ (again using \myref{eqn10} and the fact, that $\eta^p_\gamma:\FunctorT^p\gamma\to\gamma$ for $p\in \{f,h\}$).  
\par 
4. Let $u^2 := u\ast u$ be the horizontal composition. By the definition of $u^2$ we have for each $\gamma\in \CO{L}$:
\begin{equation}\label{eqn100}
u^2_\gamma = u_{\FunctorT^h\gamma}\circ \FunctorT^fu_\gamma = \FunctorT^hu_\gamma\circ u_{\FunctorT^f\gamma}.
\end{equation}
From Fig.\ \ref{Fig1,8}, we get
\begin{equation}\label{eqn101}
\pi_2\circ \o{\pi}_2 = \pi_2\circ \mu^p_\gamma
\end{equation}
where $\mu^p_\gamma = p^\ast\pi_2$. In the sequel, we use this for $p:=f$ and $p:=h$. The diagrams 
\begin{figure}[H]
\[
\xymatrix{
L\times_S(L\times_SI)\ar[dr]_{\mu^h_\gamma = h^\ast\pi'_2}
			& L\times_S(L\times_BI)\ar[l]^{\FunctorT^hu_\gamma}\ar[d]^{h^\ast\pi_2}\ar[r]^{\tilde{\pi_2}}	
						& L\times_BI\ar[d]^{\pi_2}	\\
			& L\times_SI\ar[r]_{\pi'_2}	
						& I 			\\
}\]
\caption{Compatibility with co-unit, part 1}\label{Fig8.3}
\end{figure}
and
\begin{figure}[H]
\[
\xymatrix{
{L\times_B(L\times_BI)}\ar@/_1pc/[ddr]_{(\FunctorT^f)^2\gamma}\ar@/^1pc/[drr]^{\o{\pi}_2}\ar@{ (-->}[dr]_{u_{\FunctorT^f\gamma}}	& &	\\
					& {L\times_S(L\times_BI)}\ar[r]_{\tilde{\pi}_2}\ar[d]_{\FunctorT^h\FunctorT^f\gamma}& {L\times_BI}\ar[d]^{h\circ\FunctorT^f\gamma}\\
					& L\ar[r]_{h}																& {S}					 \\
}\]
\caption{Compatibility with co-unit, part 2}\label{Fig8.2}
\end{figure}
\noindent 
commute: In the first diagram, the triangle commutes by applying $h^\ast$ to \myref{eqn10} interpreted as diagram in $\CO{S}$. The square is just the pullback which arises from pulling back $\pi_2: h\circ \FunctorT^f\gamma\to h\circ \gamma$ along $h$. We denote with $\tilde{\pi}_2$ the second projection in this case. 
\par 
The second diagram is just Figure \ref{Fig8} taken at $\FunctorT^f\gamma$ instead of $\gamma$ where the same $\tilde{\pi}_2$ occurs again. Thus 
\begin{align*}
\pi_2'\circ\mu^h_\gamma\circ u^2_\gamma 
 		&= \pi_2'\circ\mu^h_\gamma\circ \FunctorT^hu_\gamma\circ u_{\FunctorT^f\gamma}	&& \mbox{By \myref{eqn100}}			\\
		&= \pi_2\circ \tilde{\pi}_2 \circ u_{\FunctorT^f\gamma}								&& \mbox{Figure \ref{Fig8.3}}			\\
		&= \pi_2\circ \o{\pi}_2 															&& \mbox{Figure \ref{Fig8.2}}			\\
		&= \pi_2\circ \mu^f_\gamma														&& \mbox{By \myref{eqn101}} \\
		&= \pi_2'\circ u_\gamma\circ \mu^f_\gamma							&& \mbox{By \myref{eqn10}} \\
\end{align*}
On the other hand, by \myref{eqn10} and the fact that $\mu^f$ and $\mu^h$ are $\gamma$-indexed families of arrows from $(\FunctorT^f)^2\gamma$ to $(\FunctorT^f)\gamma$ and $(\FunctorT^h)^2\gamma$ to $(\FunctorT^h)\gamma$, resp., we obtain 
\[\FunctorT^h\gamma\circ u_\gamma\circ\mu^f_\gamma = \FunctorT^f\gamma\circ\mu^f_\gamma = (\FunctorT^f)^2\gamma.\]
Since $u^2$ is a $\gamma$-indexed family of arrows from $(\FunctorT^f)^2\gamma$ to $(\FunctorT^h)^2\gamma$, we also have  
\[\FunctorT^h\gamma\circ \mu^h_\gamma\circ u^2_\gamma = (\FunctorT^h)^2\gamma\circ u^2_\gamma = (\FunctorT^f)^2\gamma.\]
Because $\FunctorT^h\gamma = \pi_1'$ and $\pi_2'$ are jointly monic, the proof is complete.  \qed

\paragraph{Proof of Proposition \ref{prop4}:} 
''$\Rightarrow$'' follows immediately from \myref{eqn11.5} and Proposition \ref{prop2} applied to the coherence witness $\xi$.
\par
''$\Leftarrow$'': We call a sequence $(y_i)_{i\in\{0, 1, \ldots, m\}}$ of elements in $L$ an {\em alternating sequence} (of $a$ and $r$), if $m\in\N$ and the following conditions hold:
\begin{enumerate}
\item[a)] for all even $i\in\{0,\ldots, m-1\}: (y_{i},y_{i+1})\in ker(p)$
\item[b)] for all odd $i\in\{0,\ldots, m-1\}: (y_{i},y_{i+1})\in ker(-p)$
\end{enumerate}
where $p\in\{a,r\}$ and $-a=r$ and $-r=a$. $m+1$ is called the length of the sequence. A sequence is called {\em proper} if $y_i\not= y_j$ for all $i\in \{0, 1, \cdots, m\}$ and $j\in \{0, 1, \cdots, m-1\}$ with $i\not=j$\footnote{
	Thus, $y_0=y_m$ is allowed.
}.
\par 
For the rear pullback span with canonical descent data $\xi^\beta$, $\xi^\tau$, we define for any alternating sequence $\sigma=(y_i)_{i\in\{0, 1, \ldots, m\}}$ a bijection $\xi_\sigma: \gamma^{-1}y_0\to \gamma^{-1}y_m$ as follows: For $m=0$:  $\xi_\sigma:=id_{\gamma^{-1}y_0}$ and for $m\geq 1$
\begin{equation}\label{eqn-sequence-to-iso}
\xi_\sigma:= \xi^{\_!}_{y_{m-1},y_{m}}\circ\cdots \circ\xi^{-p!}_{y_1,y_2}\circ \xi^{p!}_{y_0,y_1}\quad\mbox{where}\; a!=\tau \;\mbox{and}\; r!=\beta.
\end{equation}
Obviously, for a domain cycle $c=(x_i)_{i\in\{0, 1, \ldots, 2k+1\}}$ 
\[\sigma_c=(x_0,x_1,\ldots,x_{2k+1},x_0)\]
is an alternating sequence, thus we can reformulate condition \myref{eqn-condition-coherence} as $\xi_{\sigma_c}=id_{\gamma^{-1}x_0}$ for all domain cycles $c$. We claim that the following conditions are equivalent:
\begin{enumerate}
  \item\label{seq1} $\xi_{\sigma_c}=id_{\gamma^{-1}x_0}$ for all domain cycles $c=(x_i)_{i\in\{0, 1, \ldots, 2k+1\}}$.
  \item\label{seq3} $\xi_\sigma=\xi_{\sigma'}$ for all alternating sequences $\sigma=(y_i)_{i\in\{0, 1, \ldots, m\}}$ and $\sigma'=(z_i)_{i\in\{0, 1, \ldots, n\}}$ with $y_0=z_0$ and $y_m=z_n$ ({\em Independence of representative on paths from $y_0$ to $y_m$}).
\end{enumerate}
Assume for the moment that this is true, then \ref{seq3}.\ is true because \ref{seq1}.\ is the assumption of the proposition. We can then use this independence of representative to uniquely construct a coherence witness, i.e.\ a family $(\xi_{e,e'})_{(e,e')\in ker(s)}$ (where $s=\bar{a}\circ r = \bar{r}\circ a$) of bijections which satisfies neutrality and associativity from Proposition \ref{prop2} and for which \myref{eqn11.5} is valid: Clearly, $(x,x')\in ker(s)$ iff there exists an alternating sequence $\sigma = (y_i)_{i\in\{0, 1, \ldots, m\}}$ with $x=y_0$ and $x'=y_m$ such that 
\[\xi_{x,x'}:=\xi_\sigma\]
does not depend on the choice of $\sigma$. Neutrality follows from \myref{eqn-sequence-to-iso} for sequences of length $0$, \myref{eqn11.5} is ensured by sequences of length $1$.
\par
To show associativity we define the composition of two alternating sequences by
\begin{itemize}
  \item $\sigma'\circ\sigma:=(y_0,\ldots,y_m=z_0,\ldots,z_n)$ if $mn=0$ or $m,n\geq1$ and $(y_{m-1},y_m)\in ker(p)$, $(z_0,z_{1})\in ker(-p)$
  \item $\sigma'\circ\sigma:=(y_0,\ldots,y_{m-1},z_1,\ldots,z_n)$ if $m,n\geq1$ and $(y_{m-1},y_m)\in ker(p)$, $(z_0,z_{1})\in ker(p)$
\end{itemize}
Again by the independence of representative we obtain for each pair $(x,x'), (x', x'')\in ker(s)$ (with representing alternating sequences $\sigma$, $\sigma'$): $\xi_{\sigma'}\circ \xi_{\sigma} = \xi_{\sigma'\circ\sigma}$, hence associativity.    
\par 
It remains to prove the equivalence ''\ref{seq1}.\ $\iff$ \ref{seq3}.''. It is easy to show that one can restrict oneself to proper alternating sequences. Then ''\ref{seq3}.\ $\Rightarrow$ \ref{seq1}.'' because \ref{seq1}.\ is a special case of \ref{seq3}.\ with $m=2k+2$ for $k\in\N$, $p=a$, and $n=0$. Thus, it remains to show ''\ref{seq1}.\ $\Rightarrow$ \ref{seq3}.''.
\par
Note first that the equation $\xi_{e,e'} = (\xi_{e',e})^{-1}$ (cf.\ Proposition \ref{prop2}) carries over to alternating sequences: If $\sigma = (y_0, \ldots, y_m)$, then $\sigma^-:= (y_m, \ldots, y_0)$ is an alternating sequence with
\begin{equation}\label{eqn5.1}\xi_\sigma = (\xi_{\sigma^-})^{-1}. \end{equation}
\par 
For the proof of ''\ref{seq1}.\ $\Rightarrow$ \ref{seq3}.'', we use induction over $n$. For $n=0$ we have $y_0=z_0=y_m$ and $\xi_{\sigma'}=id_{\gamma^{-1}z_0}$. For $m=0$ and $m=1$ we have $\xi_\sigma=id_{\gamma^{-1}y_0}=id_{\gamma^{-1}y_m}=id_{\gamma^{-1}z_0}=\xi_{\sigma'}$. Thus there remain two major cases
\begin{enumerate}
  \item $m=2k+2$ for some $k\in\N$: Then either $\sigma$ represents a domain cycle (if $p=a$) or the reverse cycle $\sigma^-$ is a domain cycle (if $p=r$). Both situations yield $\xi_\sigma = id_{\gamma^{-1}y_0} = \xi_{\sigma'}$ (cf.\ \myref{eqn5.1}).
  \item $m=2k+3$ for some $k\in\N$: If $p=a$ we know that $y_{m-1}\not= y_1$, since $\sigma$ is proper, thus the alternating sequence $\o{\sigma}=(y_{m-1},y_1,\ldots,y_{m-1})$ represents a domain cycle which is connected to $\sigma$ via $\xi_\sigma =  \xi^\tau_{y_{m-1},y_m}\circ\xi_{\o{\sigma}}\circ\xi^\tau_{y_0,y_{m-1}}$ (using associativity for $(y_0=y_m, y_{m-1}),(y_{m-1}, y_1) \in ker(a)$). By assumption
\[\xi_\sigma = \xi^\tau_{y_{m-1}, y_m} \circ \xi^\tau_{y_0, y_{m-1}} = \xi^\tau_{y_0, y_m} = id_{\gamma^{-1}y_0} = \xi_{\sigma'}\]
because $y_0 = y_m$. If $p=r$ the same argument can be carried out with $\o{\sigma} = (y_1, ..., y_{m-1}, y_1)$.
\end{enumerate}
Now we show the induction step to $n\geq 1 $ under the hypothesis that the assertion is true for all pairs $(m,n')$ with $n'<n$. Again there are several cases with possible subcases:
\begin{enumerate}
  \item $z_1=y_0$: This means $z_1=y_0=z_0$ and thus $\xi_{\sigma'}=\xi_{\sigma'_1}$ for the sequence $\sigma'_1=(z_1,\ldots,z_n)$. $\xi_\sigma=\xi_{\sigma'_1}$, however, holds by induction hypothesis.
  \item $z_1=y_k$ for some $1\leq k\leq m$: By induction hypothesis we have $\xi_{\sigma_1}=\xi_{\sigma'_1}$ and $\xi_{\sigma_2}=\xi_{\sigma'_2}$ for the subsequences $\sigma_1=(y_0,\ldots,y_k)$, $\sigma_2=(y_k,\ldots,y_m)$, $\sigma'_1=(z_0,z_1)$, $\sigma'_2=(z_1,\ldots,z_n)$ thus we also obtain $\xi_{\sigma}=\xi_{\sigma_2}\circ\xi_{\sigma_1}=\xi_{\sigma'_2}\circ\xi_{\sigma'_1}=\xi_{\sigma'}$.
  \item $z_1\not=y_k$ for all $ 0\leq k\leq m$:
  \begin{enumerate}
    \item $(y_0,y_1)\in ker(p)$, $(z_0,z_1)\in ker(p)$: Then $\sigma_1=(z_1,y_1,\ldots,y_m)$ is a proper alternating sequence. By induction hypothesis we have $\xi_{\sigma_1}=\xi_{\sigma'_1}$ for the alternating sequence $\sigma'_1=(z_1,\ldots,z_n)$, thus   $\xi_{\sigma}=\xi_{\sigma_1}\circ\xi^{p!}_{y_0,z_1}=\xi_{\sigma'_1}\circ\xi^{p!}_{z_0,z_1}=\xi_{\sigma'}$.
    \item  $(y_0,y_1)\in ker(p)$, $(z_0,z_1)\in ker(-p)$: Then $\sigma_1=(z_1,z_0=y_0,y_1,\ldots,y_m)$ is a proper alternating sequence. By induction hypothesis we have $\xi_{\sigma_1}=\xi_{\sigma'_1}$ for the alternating sequence $\sigma'_1=(z_1,\ldots,z_n)$, thus we obtain, finally, $\xi_{\sigma}=\xi_{\sigma_1}\circ\xi^{-p!}_{z_0,z_1}=\xi_{\sigma'_1}\circ\xi^{-p!}_{z_0,z_1}=\xi_{\sigma'}$. \qed
  \end{enumerate}
\end{enumerate}
\bibliographystyle{eptcsstyle/eptcs}
\bibliography{biblio/all}
\end{document}